\documentclass[12pt,prd,aps,amssymb,amsmath,tightenlines,showpacs]{revtex4}
\usepackage{psfig}

\usepackage{graphics}
\usepackage[pdftex]{graphicx}
\newcommand{\half}{\mbox{$\textstyle\frac{1}{2}$}}
\begin{document}
\preprint{}

\title{Extension of $\mathcal{PT}$-Symmetric Quantum Mechanics
to Quantum Field Theory with Cubic Interaction}

\author{Carl~M.~Bender\footnote{Permanent address: Department of Physics,
Washington University, St. Louis, MO 63130, USA.}, Dorje~C.~Brody,
and Hugh~F.~Jones}

\affiliation{Blackett Laboratory, Imperial College, London SW7
2BZ, UK}

\date{\today}

\begin{abstract}
It has recently been shown that a non-Hermitian Hamiltonian $H$
possessing an unbroken $\mathcal{PT}$ symmetry (i) has a real
spectrum that is bounded below, and (ii) defines a unitary theory
of quantum mechanics with positive norm. The proof of unitarity
requires a linear operator $\mathcal{C}$, which was originally
defined as a sum over the eigenfunctions of $H$. However, using
this definition to calculate $\mathcal{C}$ is cumbersome in
quantum mechanics and impossible in quantum field theory. An
alternative method is devised here for calculating $\mathcal{C}$
directly in terms of the operator dynamical variables of the
quantum theory. This new method is general and applies to a
variety of quantum mechanical systems having several degrees of
freedom. More importantly, this method is used to calculate the
$\mathcal{C}$ operator in quantum field theory. The $\mathcal{C}$
operator is a new time-independent observable in
$\mathcal{PT}$-symmetric quantum field theory.
\end{abstract}


\maketitle

\section{Introduction}
\label{s1}

In 1998 numerical and perturbative methods were used to establish
the surprising result that the class of non-Hermitian Hamiltonians
\begin{eqnarray}
H=p^2+x^2(ix)^\epsilon \qquad(\epsilon>0) \label{e1}
\end{eqnarray}
has a positive real spectrum~\cite{BB}. In Ref.~\cite{BB} it was
conjectured that the spectral positivity was associated with the
space-time reflection symmetry ($\mathcal{PT}$ symmetry) of the
Hamiltonian. The Hamiltonian $H$ in (\ref{e1}) is $\mathcal{PT}$
symmetric because under parity reflection $\mathcal{P}$ we have
$x\to-x$ and $p\to-p$ and under time reversal $\mathcal{ T}$ we
have $x\to x$, $p\to-p$, and $i\to-i$. Other
$\mathcal{PT}$-symmetric quantum mechanical models have been
examined~\cite{BBM,R1,R2,R3,R4}, and a proof of the positivity of
the spectrum of $H$ in (\ref{e1}) was subsequently given by Dorey
{\it et al.} \cite{DDT}.

The discovery that the spectra of many $\mathcal{PT}$-symmetric
Hamiltonians are real and positive raised a fundamental question:
Does a non-Hermitian Hamiltonian such as $H$ in (\ref{e1}) define
a consistent unitary theory of quantum mechanics, or is the
positivity of the spectrum merely an intriguing mathematical
property of special classes of complex Sturm-Liouville eigenvalue
problems? To answer this question it is necessary to know whether
the Hilbert space on which the Hamiltonian acts has an inner
product associated with a positive norm. Furthermore, it is
necessary to determine whether the dynamical time evolution
induced by such a Hamiltonian is unitary; that is, whether the
norm is preserved in time.

Recently, a definitive answer to this question was found
\cite{AM,BBJ}. For a complex non-Hermitian Hamiltonian having an
{\it unbroken} $\mathcal{PT}$ symmetry, a linear operator
$\mathcal{C}$ that commutes with both $H$ and $\mathcal{PT}$ can
be constructed. The inner product with respect to $\mathcal{ CPT}$
conjugation satisfies the requirements for the theory defined by
$H$ to have a Hilbert space with a positive norm and to be a
consistent unitary theory of quantum mechanics. [The term {\it
unbroken} ${\cal PT}$ symmetry means that every eigenfunction of
$H$ is also an eigenfunction of the ${\cal PT}$ operator. This
condition guarantees that the eigenvalues of $H$ are real. The
Hamiltonian in (\ref{e1}) has an unbroken $\mathcal{PT}$ symmetry
for all real $\epsilon \geq0$.]

We emphasize that in a conventional quantum theory the inner
product is formulated with respect to ordinary Dirac Hermitian
conjugation (complex conjugate and transpose). Unlike conventional
quantum theory, the inner product for a quantum theory defined by
a non-Hermitian $\mathcal{PT}$-symmetric Hamiltonian depends on
the Hamiltonian itself and is thus determined dynamically. One can
view this new kind of quantum theory as a ``bootstrap'' theory
because one must solve for the eigenstates of $H$ before knowing
what the Hilbert space and the associated inner product of the
theory are. The Hilbert space and inner product are then
determined by these eigenstates.

The key breakthrough in understanding these novel non-Hermitian
quantum theories was the discovery of the operator $\mathcal{C}$
\cite{BBJ}. This operator possesses three crucial properties.
First, it commutes with the space-time reflection operator
$\mathcal{PT}$,
\begin{eqnarray}
[\mathcal{C},\mathcal{PT}]=0, \label{e2}
\end{eqnarray}
although $\mathcal{C}$ does not commute with $\mathcal{P}$ or
$\mathcal{T}$ separately. Second, the square of $\mathcal{C}$ is
the identity,
\begin{eqnarray}
\mathcal{C}^2={\bf 1}, \label{e3}
\end{eqnarray}
which allows us to interpret $\mathcal{C}$ as a reflection
operator. Third, $\mathcal{C}$ commutes with $H$,
\begin{eqnarray}
[\mathcal{C},H]=0, \label{e4}
\end{eqnarray}
and thus is time independent. To summarize, $\mathcal{C}$ is a new
time-independent $\mathcal{PT}$-symmetric reflection operator.

The question now is how to construct $\mathcal{C}$ for a given
$H$. In Refs.~\cite{BBJ,BBJ2} it was shown how to express the
$\mathcal{C}$ operator in coordinate space as a sum over the
appropriately normalized eigenfunctions $\phi_n(x)$ of the
Hamiltonian $H$. These eigenfunctions satisfy
\begin{eqnarray}
H\phi_n(x)=E_n\phi_n(x), \label{e5}
\end{eqnarray}
and, without loss of generality, their overall phases are chosen
so that
\begin{eqnarray}
\mathcal{PT}\phi_n(x)=\phi_n(x). \label{e6}
\end{eqnarray}
With this choice of phase, the eigenfunctions are then normalized
according to
\begin{eqnarray}
\int_C dx\,[\phi_n(x)]^2=(-1)^n. \label{e7}
\end{eqnarray}
The contour of integration $C$ is described in detail in
Ref.~\cite{BBJ}. For the quantum mechanical theories discussed in
this paper, all of which have a cubic interaction term, the
contour $C$ can be taken to lie along the real-$x$ axis.

In terms of the eigenfunctions defined above, the statement of
completeness for a theory described by a non-Hermitian
$\mathcal{PT}$-symmetric Hamiltonian reads \cite{BBJ}
\begin{eqnarray}
\sum_n(-1)^n\phi_n(x)\phi_n(y)=\delta(x-y) \label{e8}
\end{eqnarray}
for real $x$ and $y$. The coordinate-space representation of
$\mathcal{C}$ is~\cite{BBJ}
\begin{eqnarray}
\mathcal{C}(x,y)=\sum_n\phi_n(x)\phi_n(y). \label{e9}
\end{eqnarray}
Only a {\it non-Hermitian} $\mathcal{PT}$-symmetric Hamiltonian
possesses a $\mathcal{C}$ operator distinct from the parity
operator $\mathcal{P}$. Indeed, if one evaluates the summation
(\ref{e9}) for a $\mathcal{PT}$-symmetric Hamiltonian that is also
Hermitian, the result is $\mathcal{P}$, which in coordinate space
is $\delta(x+y)$.

The coordinate-space formalism using (\ref{e9}) has been applied
successfully to
\begin{eqnarray}
H=\half p^2+\half\mu^2 x^2+i\epsilon x^3, \label{e10}
\end{eqnarray}
and $\mathcal{C}$ was constructed perturbatively to order
$\epsilon^ 3$~\cite{BMW}. This formalism has also been applied to
calculate $\mathcal{C}$ to order $\epsilon$ for the complex
H\'enon-Heiles Hamiltonian \cite{HH}
\begin{eqnarray}
H=\half\left(p_x^2+p_y^2\right)+\half\left(x^2+y^2\right)+i\epsilon
x^2y, \label{e11}
\end{eqnarray}
which has two degrees of freedom, and for the Hamiltonian
\begin{eqnarray}
H=\half\left(px^2+p_y^2+p_z^2\right)+\half\left(x^2+y^2+z^2\right)+i\epsilon
xyz, \label{e12}
\end{eqnarray}
which has three degrees of freedom \cite{BBRR}.

Calculating the operator $\mathcal{C}$ by direct evaluation of the
sum in (\ref{e9}) is difficult in quantum mechanics because it is
necessary to determine all the eigenfunctions of $H$. Such a
procedure cannot be used at all in quantum field theory because
there is no simple analog of the Schr\"odinger eigenvalue problem
(\ref{e5}) and its associated coordinate-space eigenfunctions.

In this paper we devise an elementary operator technique for
calculating $\mathcal{C}$ for the important case of quantum
theories having {\it cubic} interactions, and we demonstrate that
our new method readily generalizes from quantum mechanics to
quantum field theory. In Sec.~\ref{s2} we introduce a general
operator representation for $\mathcal{C}$ of the form $e^{Q(x,p)}
\mathcal{P}$, where $x$ and $p$ are the dynamical variables. This
representation is especially convenient for incorporating the
three requirements (\ref{e2}) -- (\ref{e4}). In Sec.~\ref{s3} we
calculate $\mathcal{C}$ to seventh order in powers of $\epsilon$
for the Hamiltonian (\ref{e10}) using this operator technique.
Then, in Sec.~\ref{s4} we calculate $\mathcal{C}$ for the
Hamiltonians (\ref{e11}) and (\ref{e12}) to order $\epsilon^3$. In
Sec.~\ref{s5} we apply operator methods to calculate $\mathcal{C}$
for the massless Hamiltonian $H=\half p^2+ix^3$. We derive
recursion relations for the operator representation of
$\mathcal{C}$ in Sec.~\ref{s6}. In Sec.~\ref{s7} we calculate
$\mathcal{C}$ to order $\epsilon^2$ for the self-interacting
scalar quantum field theory described by the Hamiltonian
\begin{eqnarray}
H=\int\!\!d^Dx\,\Big\{\half\pi^2({\bf x},t)+\half[\nabla_{\!\bf
x}\varphi({\bf x},t)]^2+\half\mu^2\varphi^2({\bf
x},t)+i\epsilon\varphi^3({\bf x},t)\Big\} \label{e13}
\end{eqnarray}
in $(D+1)$-dimensional Minkowski spacetime. In Sec.~\ref{s8} we
calculate $\mathcal{C}$ for cubic scalar quantum field theories
with interactions of the form $i\epsilon\varphi_1^2\varphi_2$ and
$i\epsilon\varphi_1\varphi_2\varphi_3$. An alternative
perturbative calculation of $\mathcal{C}$ for an $i\epsilon
\varphi^3$ quantum field theory using diagrammatic and
combinatoric methods is given in Appendix~\ref{app}. Some
concluding remarks are in Sec.~\ref{s9}.

The principal accomplishment of this paper is the derivation in
Secs.~\ref{s7} and \ref{s8} and Appendix~\ref{app} of the
$\mathcal{C}$ operator for cubic quantum field theories. Cubic
quantum field theories, such as that in (\ref{e13}), are not just
of mathematical interest. Such theories emerge in the study of
Reggeon field theory \cite{BROWER} and in the analysis of the
Lee-Yang edge singularity \cite{EDGE}. For these quantum field
theories the operator $\mathcal{C}$ is a new conserved quantity.
Knowing how to calculate this operator is crucial because it is
necessary to have $\mathcal{C}$ in order to construct observables
and to evaluate matrix elements of field operators. Our
calculation of $\mathcal{C}$ is a major step in our ongoing
program to obtain new physical models by extending conventional
quantum mechanics and quantum field theory into the complex
domain.

\section{General Form for the Operator $\mathcal{C}$}
\label{s2}

To prepare for calculating $\mathcal{C}$ we show in this section
that it is advantageous to represent $\mathcal{C}$ as a product of
the exponential of a Hermitian operator $Q$ and the parity
operator $\mathcal{P}$:
\begin{eqnarray}
\mathcal{C}=e^{Q(x,p)}\mathcal{P}. \label{e14}
\end{eqnarray}
This representation was first noticed in Ref.~\cite{BMW}.

\subsection{Previous Work on Calculating $\mathcal{C}$}

The objective of the investigation in Ref.~\cite{BMW} was to use
perturbative methods to calculate $\mathcal{C}$ for the
Hamiltonian $H=\half p^2+\half x^2+i \epsilon x^3$, where
$\epsilon$ is treated as a small parameter. In Ref.~\cite{BMW} the
operator $\mathcal{C}$ was obtained to third order in $\epsilon$
in coordinate space. The procedure was as follows: First, the
Schr\"odinger equation
\begin{eqnarray}
-\half\phi_n''(x)+\half x^2\phi_n(x)+i\epsilon
x^3\phi_n(x)=E_n\phi_n(x) \label{e15}
\end{eqnarray}
was solved for the energies $E_n$ and for the wave functions
$\phi_n(x)$ as Rayleigh-Schr\"odinger perturbation series in
powers of $\epsilon$. The series for $\phi_n(x)$ has the form
\begin{eqnarray}
\phi_n(x)=i^n\frac{a_n}{\pi^{1/4}2^{n/2}\sqrt{n!}}e^{-\frac{1}{2}x^2}
\left[H_n(x)-iA_n(x)\epsilon-B_n(x)\epsilon^2+iC_n(x)\epsilon^3+\cdots\right],
\label{e16}
\end{eqnarray}
where $H_n(x)$ is the $n$th Hermite polynomial and $A_n(x)$,
$B_n(x)$, and $C_n (x)$ are polynomials in $x$ of degree $n+3$,
$n+6$, and $n+9$, respectively. These polynomials were expressed
as linear combinations of Hermite polynomials. The value of $a_n$,
\begin{eqnarray}
a_n=1+{\textstyle\frac{1}{144}}(2n+1)(82n^2+82n+87)\epsilon^2+\mathcal{O}(
\epsilon^4), \label{e17}
\end{eqnarray}
ensures that the eigenfunctions are normalized according to
$\int_{-\infty}^{ \infty}dx\,\phi_n^2(x)=(-1)^n$, as in
(\ref{e7}). The factor $i^n$ in (\ref{e16}) is included to satisfy
the requirement in (\ref{e6}) that $\mathcal{
PT}\phi_n(x)=\phi_n(x)$.

Finally, (\ref{e16}) was substituted into (\ref{e9}) and the
summation over $n$ was performed to obtain the operator
$\mathcal{C}(x,y)$ to order $\epsilon^3$:
\begin{eqnarray}
\mathcal{C}(x,y) &=&
\Big\{1-\epsilon\left({\textstyle\frac{4}{3}}p^3-2xy p
\right)+\epsilon^2\left[{\textstyle\frac{8}{9}}p^6-{\textstyle\frac{8}{3}}xy
p^4+(2x^2y^2-12) p^2 \right]\nonumber\\
&&\quad
-\,\epsilon^3\left[{\textstyle\frac{32}{81}}p^9-{\textstyle\frac{16}{
9}}xyp^7+\left({\textstyle\frac{8}{3}}x^2y^2-{\textstyle\frac{176}{5}}\right)
p^5-\left({\textstyle\frac{4}{3}}x^3y^3-48xy\right)p^3\right.\nonumber\\
&&\quad
\left.-(8x^2y^2-64)p\right]+\mathcal{O}(\epsilon^4)\Big\}\delta(x+y),
\label{e18}
\end{eqnarray}
where $p=-i\,d/dx$. This expression is complicated, but it was
observed that it simplifies considerably when the expression in
curly brackets is rewritten in exponential form:
\begin{eqnarray}
\mathcal{C}(x,y)=e^{\epsilon Q_1+\epsilon^3
Q_3+\ldots}\delta(x+y)+\mathcal{O} (\epsilon^5). \label{e19}
\end{eqnarray}
In this form the differential operators $Q_1$ and $Q_3$ are simply
\begin{eqnarray}
Q_1 &=& -{\textstyle\frac{4}{3}} p^3-2xp x, \nonumber\\
Q_3 &=&
{\textstyle\frac{128}{15}}p^5+{\textstyle\frac{40}{3}}xp^3x+8x^2p
x^2-12\,p . \label{e20}
\end{eqnarray}

The main features of the exponential representation (\ref{e19})
are that only odd powers of $\epsilon$ appear in the exponent, the
coefficients are all real, and the derivative operators act on the
parity operator $\delta(x+y)$. Also, $e^{\epsilon Q_1+\epsilon^3
Q_3}$ is Hermitian.

\subsection{New Approach to Calculating $\mathcal{C}$}

The perturbative calculation described above suggests that a
simpler and more direct way to calculate $\mathcal{C}$ is to seek
an operator representation of it in the form
$e^{Q(x,p)}\mathcal{P}$, where $Q(x,p)$ is a Hermitian function of
the operators $x$ and $p$. We will show that $Q(x,p)$ can be found
by solving elementary operator equations and that it is not
necessary to find the eigenfunctions to determine $Q$. Thus, the
technique introduced in this paper immediately generalizes to
quantum field theory. To find the operator equations satisfied by
$Q$ we substitute $\mathcal{C}=e^Q\mathcal{P}$ into the three
equations (\ref{e2}) -- (\ref{e4}). The details are described in
Sec.~\ref{s3}.

We claim that the representation $\mathcal{C}=e^Q\mathcal{P}$ is
general. Let us illustrate this simple representation for
$\mathcal{C}$ in two elementary cases: First, consider the shifted
harmonic oscillator
\begin{eqnarray}
H=\half p^2+\half x^2+i\epsilon x. \label{e21}
\end{eqnarray}
This Hamiltonian has an unbroken $\mathcal{PT}$ symmetry for all
real $\epsilon $. Its eigenvalues
$E_n=n+\frac{1}{2}+\frac{1}{2}\epsilon^2$ are all real. The
$\mathcal{C}$ operator for this theory is given exactly by
$\mathcal{C}=e^Q \mathcal{P}$, where $Q=-\epsilon p$. Note that in
the limit $\epsilon\to0$, where the Hamiltonian becomes Hermitian,
$\mathcal{C}$ becomes identical with $\mathcal{P}$.

As a second example, consider the non-Hermitian $2\times2$ matrix
Hamiltonian
\begin{eqnarray}
H=\left(\begin{array}{cc} re^{i\theta}&s\cr s &
re^{-i\theta}\end{array}\right), \label{e22}
\end{eqnarray}
which was discussed in Ref.~\cite{BBJ2}. This Hamiltonian is
$\mathcal{PT}$ symmetric, where $\mathcal{P}$ is the Pauli matrix
$\sigma_1=\left({0~1\atop1~0} \right)$ and $\mathcal{T}$ is
complex conjugation. This Hamiltonian has an unbroken
$\mathcal{PT}$ symmetry when $s^2\geq r^2\sin^2\theta$. The
$\mathcal{C }$ operator in the unbroken region is
\begin{eqnarray}
{\cal C}=\frac{1}{\cos\alpha}\left(\begin{array}{cc} i \sin\alpha
& 1\cr 1 & -i \sin\alpha \end{array}\right), \label{e23}
\end{eqnarray}
where $\sin\alpha=(r/s)\,\sin\theta$. Our new way to express
$\mathcal{C}$ is to rewrite it in the form
$\mathcal{C}=e^Q\mathcal{P}$. Thus, the Hermitian operator $Q$ has
the form
\begin{eqnarray}
Q=\half\sigma_2\ln\left(\frac{1-\sin\alpha}{1+\sin\alpha}\right),
\label{e24}
\end{eqnarray}
where $\sigma_2=\left({0~-i\atop\!\!i~\,\,0}\right)$. Again,
observe that in the limit $\theta\to0$, where the Hamiltonian
becomes Hermitian, the $\mathcal{C}$ operator becomes identical
with $\mathcal{P}$.

Note that $\mathcal{P}$, which is given in coordinate space as
$\delta(x+y)$, can be expressed in terms of the fundamental
operators $x$ and $p$ as
\begin{eqnarray}
\mathcal{P}(x,p)=\exp\left[\half i\pi(p^2+x^2-1)\right].
\label{e25}
\end{eqnarray}
To show that the parity operator satisfies
$\mathcal{P}x\mathcal{P}^{-1}=-x$ and
$\mathcal{P}p\mathcal{P}^{-1}=-p$, we define the operator-valued
functions $f(\tau)$ and $g(\tau)$ as
\begin{eqnarray}
f(\tau)=e^{i\tau(p^2+x^2)}x\,e^{-i\tau(p^2+x^2)},\quad
g(\tau)=e^{i\tau(p^2+x^2)}p\,e^{-i\tau(p^2+x^2)}. \label{e26}
\end{eqnarray}
Differentiating $f(\tau)$ and $g(\tau)$ once gives gives
$f'(\tau)=2g(\tau)$ and $g'(\tau)=-2f(\tau)$. A second
differentiation then leads to the differential equations
$f''(\tau)=4f(\tau)$ and $g''(\tau)=4g(\tau)$. The solutions to
these equations satisfying the initial conditions $f(0)=x$ and
$g(0)=p$ are
\begin{eqnarray}
f(\tau)=x\cos(2\tau)-ip\sin(2\tau),\quad
g(\tau)=p\cos(2\tau)-ix\sin(2\tau). \label{e27}
\end{eqnarray}
Setting $\tau=\frac{1}{2}\pi$, we get $f(\tau)=-x$ and
$g(\tau)=-p$, which establishes that the operator $\mathcal{P}$
defined in (\ref{e25}) indeed has the properties of a parity
reflection operator. Specifically, $\mathcal{P}$ is a unitary
operator that generates a rotation by $\pi$ in the $(x,p)$ plane.
Another application of $\mathcal{P}$ gives a rotation by $2\pi$ in
the $(x,p)$ plane. Hence $\mathcal{P}^2=1$. This procedure
determines $\mathcal{P}$ up to an additive phase. It is
conventional to choose the phase to be $-\frac{1}{2}\pi$, as in
(\ref{e25}).

\section{Cubic Oscillator with One Degree of Freedom}
\label{s3}

Having shown that (\ref{e14}) is a natural way to represent the
operator $\mathcal{C}$, we now demonstrate how to use this {\it
ansatz} to calculate $\mathcal{C}$ for the Hamiltonian
(\ref{e10}). The procedure is to impose the three conditions
(\ref{e2}) -- (\ref{e4}) in turn on $\mathcal{C}=e^{Q(x,p)}
\mathcal{P}$ and thereby to determine the operator-valued function
$Q(x,p)$.

First, we substitute (\ref{e14}) into the condition (\ref{e2}) to
obtain
$$e^{Q(x,p)}=\mathcal{PT}e^{Q(x,p)}\mathcal{PT}=e^{Q(-x,p)},$$
from which we conclude that $Q(x,p)$ is an {\it even} function of
$x$. Second, we substitute (\ref{e14}) into the condition
(\ref{e3}) and find that
$$e^{Q(x,p)}\mathcal{P}e^{Q(x,p)}\mathcal{P}=e^{Q(x,p)}e^{Q(-x,-p)}=1,$$
which implies that $Q(x,p)=-Q(-x,-p)$. Since we already know that
$Q(x,p)$ is an even function of $x$, we conclude that it is also
an {\it odd} function of $p$.

The remaining condition (\ref{e4}) to be imposed is that the
operator $\mathcal {C}$ commutes with $H$. Substituting
$\mathcal{C}=e^{Q(x,p)}\mathcal{P}$ into (\ref{e4}), we get
$e^{Q(x,p)}[\mathcal{P},H]+[e^{Q(x,p)},H]\mathcal{P}=0$. All of
the Hamiltonians $H$ considered in this paper can be expressed in
the form $H=H_0+\epsilon H_1$, where $H_0$ is a free field theory
(harmonic oscillator) Hamiltonian that commutes with the parity
operator $\mathcal{P}$, and $H_1$ represents the interaction. For
example, for the Hamiltonian $H=\half p^2+\half \mu^2x^2+i\epsilon
x^3$, $H_0=\half p^2+\half\mu^2x^2$ and $H_1=ix^3$. Then, this
condition reads
\begin{eqnarray}
e^{Q(x,p)}[\mathcal{P},H_1]+[e^{Q(x,p)},H]\mathcal{P}=0.
\label{e28}
\end{eqnarray}
Next, we observe that if the interaction is {\it cubic}, then
$H_1$ is {\it odd} under parity reflection; that is, $H_1$ {\it
anticommutes} with $\mathcal{P}$. Hence, for quantum theories with
cubic interaction Hamiltonians, (\ref{e28}) reduces to
\begin{eqnarray}
\epsilon e^{Q(x,p)}H_1=[e^{Q(x,p)},H]. \label{e29}
\end{eqnarray}

We note that the structure in (\ref{e19}) is quite general; for
all cubic Hamiltonians, $Q(x,p)$ may be expanded as a series in
odd powers of $\epsilon$:
\begin{eqnarray}
Q(x,p)=\epsilon Q_1(x,p)+\epsilon^3
Q_3(x,p)+\epsilon^5Q_5(x,p)+\ldots\,. \label{e30}
\end{eqnarray}
In quantum field theory we will interpret the series coefficients
$Q_{2n+1}$ as interaction vertices (form factors) of $2n+3$ powers
of the quantum fields.

Substituting the expansion in (\ref{e30}) into the exponential
$e^{Q(x,p)}$, we get
\begin{eqnarray}
e^{Q(x,p)}\equiv
R(x,p)=1+R_1(x,p)\epsilon+R_2(x,p)\epsilon^2+R_3(x,p)\epsilon^3
+R_4(x,p)\epsilon^4+\ldots\,, \label{e31}
\end{eqnarray}
where
\begin{eqnarray}
R_1 &=& Q_1,\nonumber\\
R_2 &=& \half Q_1^2,\nonumber\\
R_3 &=& Q_3+{\textstyle\frac{1}{6}}Q_1^3,\nonumber\\
R_4 &=& \half\{Q_1,Q_3\}+{\textstyle\frac{1}{24}}Q_1^4,\nonumber\\
R_5 &=&
Q_5+{\textstyle\frac{1}{6}}(\{Q_1^2,Q_3\}+Q_1Q_3Q_1)+{\textstyle\frac{1}
{120}}Q_1^5,\nonumber\\
R_6 &=&
\half(Q_3^2+\{Q_1,Q_5\})+{\textstyle\frac{1}{24}}(\{Q_1^3,Q_3\}
+\{Q_1,Q_1Q_3Q_1\})+{\textstyle\frac{1}{720}}Q_1^6, \label{e32}
\end{eqnarray}
and so on. Here, $\{X,Y\}=XY+YX$ denotes the anticommutator.

We now substitute (\ref{e32}) into (\ref{e29}), collect the
coefficients of like powers of $\epsilon^n$ for $n=1,2,3,\ldots$,
and obtain a sequence of equations of the general form
\begin{eqnarray}
\epsilon^n:~~ \left[H_0,R_n\right]=-\{H_1,R_{n-1}\}\quad(n\geq1),
\label{e33}
\end{eqnarray}
where $R_0\equiv1$.

The equations in (\ref{e33}) can be solved systematically for the
operator-valued functions $Q_n(x,p)$ $(n=1,3,5,\ldots)$ subject to
the symmetry constraints that ensure the conditions (\ref{e2}) and
(\ref{e3}). Note that the coefficients of even powers of
$\epsilon$ contain no additional information because the equation
arising from the coefficient of $\epsilon^{2n}$ can be derived
from the equations arising from the coefficients of
$\epsilon^{2n-1},\, \epsilon^{2n-3},\,\ldots,\,\epsilon$. This
observation leads to a more effective way to express the
conditions in (\ref{e33}). The first four equations read
\begin{eqnarray}
\left[H_0,Q_1\right] &=& -2H_1,\nonumber\\
\left[H_0,Q_3\right] &=& -{\textstyle\frac{1}{6}}[Q_1,[Q_1,H_1]],\nonumber\\
\left[H_0,Q_5\right] &=&
{\textstyle\frac{1}{360}}[Q_1,[Q_1,[Q_1,[Q_1,H_1]]]]
-{\textstyle\frac{1}{6}}\left([Q_1,[Q_3,H_1]]+[Q_3,[Q_1,H_1]]\right),
\nonumber\\
\left[H_0,Q_7\right] &=&
{\textstyle\frac{1}{15120}}[Q_1,[Q_1,[Q_1,[Q_1,[Q_1,[
Q_1,H_1]]]]]]\nonumber\\
&&\,
-{\textstyle\frac{1}{360}}\left([Q_1,[Q_1,[Q_1,[Q_3,H_1]]]]+[Q_1,[Q_1[
Q_3,[Q_1,H_1]]]]\right.\nonumber\\
&& \,
\left.+[Q_1,[Q_3,[Q_1,[Q_1,H_1]]]]+[Q_3,[Q_1[Q_1,[Q_1,H_1]]]]\right)
\nonumber\\
&&
\,+{\textstyle\frac{1}{6}}\left([Q_1,[Q_5,H_1]]+[Q_5,[Q_1,H_1]]\right)
{\textstyle\frac{1}{6}}[Q_3,[Q_3,H_1]]. \label{e34}
\end{eqnarray}

We now show how to solve these equations for the Hamiltonian in
(\ref{e10}), for which $H_0=\frac{1}{2}p^2+\frac{1}{2}\mu^2x^2$
and $H_1=ix^3$. The procedure is to substitute the most general
polynomial form for $Q_n$ using arbitrary coefficients and then to
solve for these coefficients. For example, to solve the first of
the equations in (\ref{e34}), $\left[H_0,Q_1\right]=-2ix^3$, we
take as an {\it ansatz} for $Q_1$ the most general Hermitian cubic
polynomial that is even in $x$ and odd in $p$:
\begin{eqnarray}
Q_1(x,p)=Mp^3+Nxpx, \label{e35}
\end{eqnarray}
where $M$ and $N$ are undetermined coefficients. The operator
equation for $Q_1$ is satisfied if
\begin{eqnarray}
M=-{\textstyle\frac{4}{3}}\mu^{-4}\quad{\rm and}\quad
N=-2\mu^{-2}. \label{e36}
\end{eqnarray}

It is straightforward, though somewhat tedious, to continue this
process. In order to present the solutions for $Q_n(x,p)$ ($n>1$),
it is convenient to introduce the following notation: Let
$S_{m,n}$ represent the {\it totally symmetrized} sum over all
terms containing $m$ factors of $p$ and $n$ factors of $x$. For
example,
\begin{eqnarray}
S_{0,0}&=&1,\nonumber\\
S_{0,3}&=&x^3,\nonumber\\
S_{1,1}&=&\half\left(xp+px\right),\nonumber\\
S_{1,2}&=&{\textstyle\frac{1}{3}}\left(x^2p+xpx+px^2\right),\nonumber\\
S_{3,1}&=&{\textstyle\frac{1}{4}}\left(xp^3+pxp^2+p^2xp+p^3x\right),\nonumber\\
S_{2,2}&=&{\textstyle\frac{1}{6}}\left(p^2x^2+x^2p^2+pxpx+xpxp+px^2p+xp^2x
\right). \label{e37}
\end{eqnarray}
The properties of the operators $S_{m,n}$ are summarized in
Ref.~\cite{BD}. One useful property is that $S_{m,n}$ can be
expressed in Weyl-ordered form in two ways:
\begin{eqnarray}
S_{m,n}=\frac{1}{2^n}\sum_{k=0}^n\left({n\atop
k}\right)x^kp^mx^{n-k}=\frac{1} {2^m}\sum_{k=0}^m\left({m\atop
k}\right) p^k x^n p^{m-k}. \label{e38}
\end{eqnarray}

We have solved the equations in (\ref{e34}) and have found $Q_1$,
$Q_3$, $Q_5$, and $Q_7$ in closed form. In terms of the
symmetrized operators $S_{m,n}$ the functions $Q_n$ are
\begin{eqnarray}
Q_1 \!&=&\!
-{\textstyle\frac{4}{3}}\mu^{-4}p^3-2\mu^{-2}S_{1,2},\nonumber\\
Q_3 \!&=&\!
{\textstyle\frac{128}{15}}\mu^{-10}p^5+{\textstyle\frac{40}{3}}
\mu^{-8}S_{3,2}+8\mu^{-6}S_{1,4}-12\mu^{-8}p,\nonumber\\
Q_5 \!&=&\!
-{\textstyle\frac{320}{3}}\mu^{-16}p^7-{\textstyle\frac{544}{3}}
\mu^{-14}S_{5,2}-{\textstyle\frac{512}{3}}\mu^{-12}S_{3,4}\nonumber\\
&&\quad-64\mu^{-10}S_{1,6}+{\textstyle\frac{24\,736}{45}}\mu^{-14}p^3
+ {\textstyle\frac{6\,368}{15}}\mu^{-12}S_{1,2},\nonumber\\
Q_7 \!&=&\!
{\textstyle\frac{553\,984}{315}}\mu^{-22}p^9+{\textstyle\frac{97\,
792}{35}}\mu^{-20}S_{7,2}+{\textstyle\frac{377\,344}{105}}\mu^{-18}S_{5,4}
\nonumber\\
&&\quad+{\textstyle\frac{721\,024}{315}}\mu^{-16}S_{3,6}+{\textstyle
\frac{1\,792}{3}}\mu^{-14}S_{1,8}-{\textstyle\frac{2\,209\,024}{105}}
\mu^{-20}p^5 \nonumber\\
&&\quad-{\textstyle\frac{2\,875\,648}{105}}\mu^{-18}S_{3,2}-{\textstyle
\frac{390\,336}{35}}\mu^{-16}S_{1,4}+{\textstyle\frac{46\,976}{5}}\mu^{-18}p.
\label{e39}
\end{eqnarray}
Combining (\ref{e14}), (\ref{e30}), and (\ref{e39}), we obtain an
explicit perturbative expansion of $\mathcal{C}$ in terms of the
fundamental operators $x$ and $p$, correct to order $\epsilon^7$.

To summarize, using the {\it ansatz} (\ref{e14}) we are able to
calculate the $\mathcal{C}$ operator to very high order in
perturbation theory. We are able to perform this calculation
because this {\it ansatz} obviates the necessity of calculating
the wave functions $\phi_n(x)$. The calculation bears a strong
resemblance to WKB theory. The {\it ansatz} used in performing a
semiclassical calculation is also an exponential of a power
series. The advantage of using WKB to calculate the energy
eigenvalues is that {\it to all orders} in powers of $\hbar$ it is
possible to construct a system of equations like those in
(\ref{e34}) that determine the energies, and it is never necessary
to calculate the wave function~\cite{BO}. Furthermore, only the
even terms in the WKB series are needed to determine the energy
eigenvalues. The odd terms in the series drop out of the
calculation and provide no information about the
eigenvalues~\cite{BO}. The difference between a conventional WKB
series and the series representation for $Q$ is that the first
term in a WKB series is proportional to $\hbar^{-1}$ while the
series expansion for $Q(x,p)$ contains only positive powers of
$\epsilon$. Based on the results in Ref.~\cite{BMW}, however, we
believe that for a $\mathcal{PT}$-symmetric $-\epsilon x^4$
theory, the first term in the expansion of $Q(x,p)$ is
proportional to $\epsilon^{-1}$. We plan to discuss quartic
$\mathcal{PT}$-symmetric theories in a future paper.

\section{Cubic Oscillators with Several Degrees of Freedom}
\label{s4}

In this section we extend the operator techniques used in
Sec.~\ref{s3} to systems having two and three dynamical degrees of
freedom. Specifically, we generalize the perturbative procedure
for calculating the $\mathcal{C}$ operator for the Hamiltonian in
(\ref{e10}) and use it to calculate $\mathcal{C}$ for the
Hamiltonians in (\ref{e11}) and (\ref{e12}).

Let us first consider $H$ in (\ref{e11}), which has two degrees of
freedom. We write this Hamiltonian in the form $H=H_0+\epsilon
H_1$, where $H_0=\half p^2 +\half q^2+\half x^2+\half y^2$ and
$H_1=ix^2y$. For these operators we need to solve the system of
equations in (\ref{e34}) for the unknown operators $Q_1$, $Q_3$,
and so on. To simplify the calculation we generalize slightly the
notation in (\ref{e37}) for totally symmetric operators; to wit,
we continue to use $S_{m,n}$ to represent a totally symmetric
product of $m$ factors of $p$ and $n$ factors of $x$ but we use
$T_{m,n}$ to represent a totally symmetric product of $m$ factors
of $q$ and $n$ factors of $y$. For example,
\begin{eqnarray}
T_{1,1}&=&\half(qy+yq),\nonumber\\
T_{1,2}&=&{\textstyle\frac{1}{3}}\left(y^2q+yqy+qy^2\right)=yqy,\nonumber\\
T_{3,1}&=&{\textstyle\frac{1}{4}}\left(yq^3+qyq^2+q^2yq+q^3y\right),\nonumber\\
T_{2,1}&=&{\textstyle\frac{1}{3}}\left(q^2y+qyq+yq^2\right)=qyq.
\label{e40}
\end{eqnarray}

To solve $\left[H_0,Q_1\right]=-2H_1$, the first equation in
(\ref{e34}), we seek a Hermitian cubic polynomial in the variables
$x$, $y$, $p$, and $q$. This polynomial must be even in the
coordinate variables and odd in the momentum variables, and it
must be able to yield $H_1$ when commuted with $H_0$. We therefore
introduce the {\it ansatz}
\begin{eqnarray}
Q_1(x,y,p,q)=Mp^2q+N_1S_{1,1}y+N_2x^2q. \label{e41}
\end{eqnarray}
We substitute this {\it ansatz} into the commutator and determine
the unknown constants $M$, $N_1$, and $N_2$ by solving three
linear equations. The result is
\begin{eqnarray}
M=-{\textstyle\frac{4}{3}},\quad
N_1=-{\textstyle\frac{2}{3}},\quad N_2=-{\textstyle\frac{2}{3}}.
\label{e42}
\end{eqnarray}

Next, we turn to the second of the equations in (\ref{e34}),
$\left[H_0,Q_3 \right]=-{\textstyle\frac{1}{6}}[Q_1,[Q_1,H_1]]$,
and evaluate its right side. The resulting equation for $Q_3$ then
reads
\begin{eqnarray}
\left[H_0,Q_3\right] &=& i{\textstyle\frac{8}{27}}
\left(4x^4y+4x^2y^3+8p^2T_{2,1}-4x^2T_{2,1}+4S_{1,3}q\right.
\nonumber \\ &&\,\left.+8S_{1,1}T_{1,2}+8S_{3,1}q-3y\right).
\label{e43}
\end{eqnarray}
We now must construct the most general Hermitian fifth-degree
polynomial in the variables $x$, $y$, $p$, and $q$ that is even in
the coordinate variables, odd in the momentum variables, and has
the terms needed to produce the right side of this commutation
relation:
\begin{eqnarray}
Q_3(x,y,p,q) &=& a_1p^2q^3+a_2p^4q+a_3S_{1,1}T_{2,1}+a_4p^2T_{1,2}
+a_5 S_{3,1}y +a_6 S_{2,2}q\nonumber\\
&&\,+a_7x^2q^3+a_8x^2T_{1,2}+a_9S_{1,1}y^3+a_{10}
S_{1,3}y+a_{11}x^4q+a_{12}q. \label{e44}
\end{eqnarray}
Substituting this {\it ansatz} into (\ref{e43}), we obtain twelve
simultaneous linear equations for the unknown coefficients $a_n$,
whose solution is
\begin{eqnarray}
&& a_1={\textstyle\frac{512}{405}},\quad
a_2={\textstyle\frac{512}{405}},\quad
a_3={\textstyle\frac{1088}{405}},\quad
a_4=-{\textstyle\frac{256}{405}},\quad
a_5={\textstyle\frac{512}{405}},\quad
a_6={\textstyle\frac{288}{405}},
\nonumber\\
&& a_7=-{\textstyle\frac{32}{405}},\quad
a_8={\textstyle\frac{736}{405}},\quad
a_9=-{\textstyle\frac{256}{405}},\quad
a_{10}={\textstyle\frac{608}{405}},\quad
a_{11}=-{\textstyle\frac{128}{405}},\quad
a_{12}=-{\textstyle\frac{8}{9}}. \label{e45}
\end{eqnarray}
This completes the calculation of the operator $\mathcal{C}$ to
third order in $\epsilon$ for $H$ in (\ref{e11}).

The attractive feature of the calculational procedure described
here is that it is utterly routine and works in every order of
perturbation theory. In contrast, the technique used in
Ref.~\cite{BBRR} to calculate $\mathcal{C}$ becomes hopelessly
difficult beyond first order in powers of $\epsilon$ because the
technique used earlier requires that one calculate all of the
energy eigenstates $\phi_n(x,y)$ perturbatively for all $n$. These
eigenstates must then be substituted directly into the summation
in (\ref{e9}) that defines the operator $\mathcal{C}$. This
calculation is difficult because beyond leading order in
$\epsilon$ one encounters the challenging problems associated with
degenerate energy levels. (There are no degenerate energy levels
for Hamiltonians having just one degree of freedom.) Of course,
for any given $n$, there is a well-defined procedure for
calculating the eigenstate to any order in powers of $\epsilon$.
However, this procedure depends on the value of $n$ and, as a
result, the calculation becomes extremely complicated. The new
method of calculation presented here works because it is no longer
necessary to calculate the eigenfunctions. Thus, the difficulties
associated with degeneracy are circumvented.

We turn next to the case of a cubic oscillator having three
degrees of freedom. We express the Hamiltonian in (\ref{e12}) in
the form $H=H_0+\epsilon H_1$, where $H_0=\half p^2+\half
q^2+\half r^2+\half x^2+\half y^2+\half z^2$ represents a harmonic
oscillator Hamiltonian having three degrees of freedom and
$H_1=ixyz$ is a non-Hermitian $\mathcal{PT}$-symmetric interaction
term.

To solve $\left[H_0,Q_1\right]=-2 H_1$, the first of the equations
in (\ref{e34}), we must construct the most general Hermitian cubic
polynomial in the variables $x$, $y$, $z$, $p$, $q$, and $r$ that
is even in the coordinate variables, odd in the momentum
variables, and has the terms needed to yield $H_1$ on the right
side of this commutation relation:
\begin{eqnarray}
Q_1(x,y,z,p,q,r)=Mpqr+N(yzp+xzq+xyr). \label{e46}
\end{eqnarray}
We then substitute this {\it ansatz} into the commutator and
determine the unknown constants $M$ and $N$ by solving two linear
equations. The result is
$$M=-{\textstyle\frac{4}{3}}\quad{\rm and}\quad N=-
{\textstyle\frac{2}{3}}.$$

To solve the second of the equations in (\ref{e34}),
$\left[H_0,Q_3\right]=- {\textstyle\frac{1}{6}}[Q_1,[Q_1,H_1]]$,
we evaluate its right side. The resulting equation for $Q_3$ then
reads
\begin{eqnarray}
\left[H_0,Q_3\right] &=&
i{\textstyle\frac{8}{27}}\Big[xyz\left(x^2+y^2+z^2
\right)+(pz+rx)yqy+(py+qx)zrz+(qz+ry)xpx\nonumber\\
&&\,-(pxpyz+qyqxz+rzrxy)+2(rqpxp+rpqyq+pqrzr)\Big]. \label{e47}
\end{eqnarray}
The most general Hermitian fifth-degree polynomial in the
variables $x$, $y$, $z$, $p$, $q$, and $r$ that is even in the
coordinate variables, odd in the momentum variables, and has the
terms needed to produce the right side of this commutation
relation is:
\begin{eqnarray}
Q_3(x,y,z,p,q,r) &=& d_1\left(p^3qr+q^3pr+r^3qp\right)
+d_2[pxp(yr+zq)+qyq(xr+zp)\nonumber\\
&&\,+rzr(xq+yp)]+d_3(xpxqr+ypypr+zrzpq)+d_4(xpxyz+yqyxz
\nonumber\\ &&\,+zrzxy) + d_5 \left[ x^3(yr+zq)+
y^3(xr+zp)+z^3(xq+yp)\right] \nonumber \\ &&\,+d_6\left(
p^3yz+q^3xz+r^3xy\right). \label{e48}
\end{eqnarray}
Substituting this {\it ansatz} into (\ref{e43}), we obtain six
simultaneous equations for the unknown coefficients $d_n$, whose
solution is
\begin{eqnarray}
d_1={\textstyle\frac{128}{405}},\quad
d_2={\textstyle\frac{136}{405}},\quad
d_3=-{\textstyle\frac{64}{405}},\quad
d_4={\textstyle\frac{184}{405}},\quad
d_5=-{\textstyle\frac{32}{405}},\quad
d_6=-{\textstyle\frac{8}{405}}. \label{e49}
\end{eqnarray}
This completes the calculation of the operator $\mathcal{C}$ to
third order in $\epsilon$ for $H$ in (\ref{e12}). This calculation
is simpler than that for the Hamiltonian in (\ref{e11}) because
there is symmetry under the interchange of pairs of dynamical
variables, such as $(x,p)\leftrightarrow(y,q)$.

\section{Representation of $\mathcal{C}$ for the Massless Case}
\label{s5}

In this section we examine the massless (strong-coupling) limit of
the operator $\mathcal{C}$ for $H$ in (\ref{e10}). The massless
limit $\mu\to0$ of the massive theory is especially interesting
because this limiting case is {\it singular.} We will see that as
the mass parameter $\mu$ tends to zero, the perturbation series
representation for $Q$ in $\mathcal{C}=e^Q\mathcal{P}$ ceases to
exist and an entirely new nonpolynomial representation for $Q$
emerges.

Negative powers of $\mu$ in (\ref{e39}) are required for
dimensional consistency. As a result, each of the terms in these
perturbation series coefficients becomes singular in the massless
limit $\mu\to0$. (Note that the dimensionless perturbation
expansion parameter is $\epsilon\mu^{-5/2}$, and thus the massless
limit of the theory is equivalent to the strong-coupling limit
$\epsilon\to\infty$.)

To find the $\mathcal{C}$ operator for the massless theory it is
necessary to return to the sequence of operator equations in
(\ref{e34}) and to look for new solutions for the special case in
which $H_0=\half p^2$. The first of these equations reads
\begin{eqnarray}
\left[\half p^2,Q_1\right]=-2ix^3. \label{e50}
\end{eqnarray}
However, an examination of (\ref{e50}) reveals that it is no
longer possible to find a solution in the form of a polynomial in
the operators $p$ and $x$. The situation here is quite similar to
that considered in Ref.~\cite{BD}, in which the objective was to
calculate the time operator in quantum mechanics. In the case of
the time operator it was shown that one must generalize the
symmetric operators $S_{m,n}$ from the positive integers to the
negative integers. Specifically, if $m$ is nonnegative, then $n$
may be negative, and if $n$ is nonnegative, then $m$ may be
negative. (It is not possible for {\it both} $m$ and $n$ to be
negative.) We can display these generalized symmetric operators in
Weyl-ordered form [see (\ref{e38})]. For example,
\begin{eqnarray}
S_{-1,1} &=&
{\textstyle\frac{1}{2}\left(x\frac{1}{p}+\frac{1}{p}x\right)},
\nonumber\\
S_{-3,0} &=& {\textstyle\frac{1}{p^3}},\nonumber\\
S_{-2,2} &=&
{\textstyle\frac{1}{4}\left(\frac{1}{p^2}x^2+2x\frac{1}{p^2}x
+x^2\frac{1}{p^2}\right)},\nonumber\\
S_{-2,3} &=&
{\textstyle\frac{1}{8}\left(x^3\frac{1}{p^2}+3x^2\frac{1}{p^2}x+3x
\frac{1}{p^2}x^2+\frac{1}{p^2}x^3\right)}. \label{e51}
\end{eqnarray}

An exact, dimensionally consistent operator solution to
(\ref{e50}) is
\begin{eqnarray}
Q_1=\half{\textstyle S_{-1,4}+\alpha
S_{-5,0}=\frac{1}{32}\left(x^4\frac{1}{p}
+4x^3\frac{1}{p}x+6x^2\frac{1}{p}x^2+4x\frac{1}{p}x^3+
\frac{1}{p}x^4\right)+ \alpha\frac{1}{p^5}}, \label{e52}
\end{eqnarray}
where $\alpha$ is an arbitrary number. This solution has the
required symmetry properties; to wit, it is odd in $p$ and even in
$x$. Also, it has the same dimensions as $Q_1$ in (\ref{e39}).

The solution to the second operator equation in (\ref{e34}),
$\left[\half p^2,
Q_3\right]=-{\textstyle\frac{1}{6}}[Q_1,[Q_1,ix^3]]$, is
\begin{eqnarray}
Q_3&=&{\textstyle\frac{1}{40}}S_{-5,10}-
{\textstyle\frac{3}{32}}S_{-7,8}+\left(
{\textstyle\frac{7}{16}}+20\alpha\right)S_{-9,6}+
\left({\textstyle\frac{3}{32}}
+{\textstyle\frac{305}{8}}\alpha\right)S_{-11,4} \nonumber\\ &&
\quad+\left(-{\textstyle\frac{135}{16}}-
{\textstyle\frac{5773}{8}}\alpha +
{\textstyle\frac{75}{12}}\alpha^2\right) S_{-13,2}+\beta
S_{-15,0}, \label{e53}
\end{eqnarray}
where $\beta$ is a new arbitrary constant. Again, observe that
this solution exhibits the required symmetry properties.

A notable feature of the solutions for $Q_1$ and $Q_3$ in
(\ref{e52}) and (\ref{e53}) is that they are {\it not} unique.
Each of these solutions contains an arbitrary constant multiplying
a negative odd-integer power of $p$. There is no obvious way to
determine the values of the constants $\alpha$ and $\beta$. These
terms arise because in the massless case the Hamiltonian $H_0$ is
a function of $p$ only. In general, one can add an arbitrary
multiple of $p^{-10n- 5}$ to the solution for $Q_{2n+1}$ because
it is odd in $p$ and is dimensionally consistent. In the massive
case, where $H_0=\half p^2+\half\mu^2x^2$, there is no such
ambiguity because adding an arbitrary function of $H_0$ to $Q_n$
would violate the symmetry requirement that $Q_n$ be odd in $p$.

\section{Product Representation of $\mathcal{C}$ and Derivation of
Recursion Relations} \label{s6}

In this section we investigate the {\it product} representation of
the operator $\mathcal{C}$ that was defined in (\ref{e31}); namely
$\mathcal{C}(x,p)=R(x,p) \mathcal{P}$. At first, it may not seem
worthwhile to reconsider the product representation because it
lacks the advantages of the {\it exponential} representation
$\mathcal{C}=e^Q\mathcal{P}$ introduced in (\ref{e14}). Recall
that we argued in Sec.~\ref{s2} that the exponential
representation is convenient because it incorporates the
requirements (\ref{e2}) and (\ref{e3}) as elementary symmetry
conditions on $Q(x,p)$: $Q(x,p)=Q(-x,p)$ and $Q(x,p)=-Q(x ,-p)$.
Furthermore, we showed that the exponential representation of
$\mathcal{ C}$ in (\ref{e19}) and (\ref{e20}) is much simpler than
the product representation $\mathcal{C}(x,p)=R(x,p)\mathcal{P}$ in
(\ref{e18}). However, as we demonstrate here, the product
representation has the advantage that it can be used to construct
a recursive formula for the perturbation coefficients.

The function $R(x,p)$ in the product representation
$\mathcal{C}(x,p)=R(x,p) \mathcal{P}$ incorporates the requirement
in (\ref{e2}) as
\begin{eqnarray}
R(x,p)=R(-x,p). \label{e54}
\end{eqnarray}
Thus, $R(x,p)$ is an even function of $x$. However, the
requirement in (\ref{e3}) translates into a complicated {\it
nonlinear} condition on $R$:
\begin{eqnarray}
R(x,p)R(x,-p)=1. \label{e55}
\end{eqnarray}
We will return to this condition later.

The advantage of the product representation is that it translates
the requirement in (\ref{e4}) into a {\it linear} difference
equation. To obtain this difference equation we use the operators
$S_{m,n}$ in (\ref{e37}). (Recall that $S_{m,n}$ is a totally
symmetric combination of products of $m$ factors of $p$ and $n$
factors of $x$.) It was shown in Ref.~\cite{BD} that the operators
$S_{m,n}$ are {\it complete} in the sense that any operator may be
represented as a linear combination of these symmetric operators.
This allows us to represent $R(x,p)$ as the infinite linear
combination
\begin{eqnarray}
R(x,p)=\sum_m\sum_n\alpha_{m,n}S_{m,n}, \label{e56}
\end{eqnarray}
where $\alpha_{m,n}$ are numerical coefficients to be determined.
Substituting $\mathcal{C}(x, p)=R(x,p)\mathcal{P}$ into (\ref{e4})
then gives the condition
\begin{eqnarray}
\left[R,H_0\right]=\epsilon\left\{R,H_1\right\}. \label{e57}
\end{eqnarray}
We now substitute $R$ in (\ref{e56}) into the condition
(\ref{e57}) and use the commutation and anti-commutation relations
\cite{BD}
\begin{eqnarray}
\left[S_{m,n},x^2\right] &=& -2imS_{m-1,n+1},\nonumber\\
\left[S_{m,n},p^2\right] &=& 2inS_{m+1,n-1},\nonumber\\
\left\{S_{m,n},x^3\right\} &=&
-{\textstyle\frac{3}{2}}m(m-1)S_{m-1,n+1}. \label{e58}
\end{eqnarray}
For the Hamiltonian in (\ref{e10}) we obtain the linear recursion
relation
\begin{eqnarray}
n\,\alpha_{m-2,n}-\mu^2\,m\,\alpha_{m,n-2}=\epsilon\left[-
{\textstyle\frac{3} {2}}m(m+1)\,\alpha_{m+1,n-2}+2
\alpha_{m-1,n-4}\right]. \label{e59}
\end{eqnarray}

The boundary conditions on this partial difference equation must
be chosen so that the nonlinear constraint (\ref{e55}) is
satisfied. In the massive case $\alpha_{0,0}=1$ and $\alpha_{m,n}$
vanishes if either $m<0$ or $n<0$. In the massless case we again
have $\alpha_{0,0}=1$, but now $\alpha_{m,n}$ vanishes if either
$m>0$ or $n<0$.

One approach to solving this equation is to introduce a generating
function $g(s,t)$:
\begin{eqnarray}
g(s,t)\equiv\sum_m\sum_n\alpha_{m,n}s^m t^n. \label{e60}
\end{eqnarray}
For the massive case [the Hamiltonian in (\ref{e10}) with
$\mu\neq0$] the summation is taken over nonnegative values of $m$
and $n$. However, for the massless case ($\mu=0$) the summation
must be taken over nonpositive values of $m$ and nonnegative
values of $n$.

We then multiply (\ref{e59}) by $s^{m-1}t^{m-1}$ and rewrite the
result in the form
\begin{eqnarray}
&& s\frac{\partial}{\partial
t}\left(\alpha_{m-2,n}s^{m-2}t^n\right)-\mu^2 t
\frac{\partial}{\partial s}\left(\alpha_{m,n-2}s^mt^{n-2}\right)
\nonumber \\ &&\qquad\qquad= \epsilon\left[
-{\textstyle\frac{3}{2}}t\frac{\partial^2} {\partial s^2}
\left(\alpha_{m+1,n-2}s^{m+1}t^{n-2}\right)+2t^3 \alpha_{m-1,n-4}
s^{m-1} t^{n-4}\right]. \label{e61}
\end{eqnarray}
Summing over $m$ and $n$ and using (\ref{e60}), we obtain the
partial differential equation
\begin{eqnarray}
sg_t-\mu^2
tg_s=\epsilon\left(-{\textstyle\frac{3}{2}}tg_{ss}+2t^3g\right),
\label{e62}
\end{eqnarray}
where subscripts indicate partial differentiation. A Fourier
transform from the $s$ variable to the $r$ variable converts this
differential equation into the {\it Goursat problem} (recall that
a {\it Goursat problem} involves a wave equation written in
light-cone variables)
\begin{eqnarray}
\tilde g_{rt}=\left(\mu^2rt +{\textstyle\frac{3}{2}}i\epsilon
r^2t+2i\epsilon t^3\right)\tilde g. \label{e63}
\end{eqnarray}
This is an extremely interesting equation that merits further
study. We plan to give a detailed analysis of this partial
differential equation and of the partial difference equation
(\ref{e59}) in a future paper.

\section{Scalar Quantum Field Theory with Cubic Self-Interaction}
\label{s7}

This section extends the operator techniques introduced in
Sec.~\ref{s3} to quantum field theory. Consider the quantum field
theory described by the Hamiltonian (\ref{e13}) in
$(D+1)$-dimensional Minkowski space-time. This Hamiltonian has the
form $H=H_0+\epsilon H_1$, where
\begin{eqnarray}
H_0=\int\!\!d^Dx\left\{\half\pi^2({\bf x},t)+\half[\nabla_{\!\bf
x}\varphi ({\bf x},t)]^2+\half\mu^2\varphi^2({\bf
x},t)\right\},\quad H_1=i\!\int\!\!d^Dx\,\varphi^3({\bf x},t).
\label{e64}
\end{eqnarray}
The integrals above are performed in the spatial variable ${\bf
x}$, which lies in ${\mathbb R}^D$. In the following we use $\int
d{\bf x}=\int d^Dx$ to represent the integration in ${\mathbb
R}^D$. The field variables satisfy the equal-time canonical
commutation relation $\left[\varphi({\bf x},t),\pi({\bf y},
t)\right]=i\delta({\bf x}-{\bf y})$.

The parity operator is given formally by
$\mathcal{P}=\exp\left(\half i\pi\int d{\bf
x}\,\left[\varphi^2({\bf x},t)+\pi^2({\bf x},t)-1\right]\right)$.
As in quantum mechanics, where the operators $x$ and $p$ change
sign under parity reflection, we assume that the fields are {\it
pseudoscalars} and that they also change sign under $\mathcal{P}$:
\begin{eqnarray}
\mathcal{P}\varphi({\bf x},t)\mathcal{P}=-\varphi(-{\bf
x},t),\quad \mathcal{P}\pi({\bf x},t)\mathcal{P}=-\pi(-{\bf x},t).
\label{e65}
\end{eqnarray}

Following the approach in Sec.~\ref{s3}, we express $\mathcal{C}$
in the form $\mathcal{C}=e^{\epsilon
Q_1+\epsilon^3Q_3+\ldots}\mathcal{P}$, where $Q_{2n+1}$
($n=0,\,1,\,2,\,\ldots$) are real functionals of the field
variables $\varphi( {\bf x},t)$ and $\pi({\bf x},t)$. To find
$Q_1$ it is necessary to solve the first of the operator equations
in (\ref{e34}):
\begin{eqnarray}
\left[\int\!\!d{\bf x}\,\Big(\half\pi^2({\bf
x},t)+\half\mu^2\varphi^2({\bf x}, t)-\half\varphi({\bf
x},t)\nabla_{\!\bf x}^2\varphi({\bf x},t)\Big),Q_1\right]=
-2i\int\!\!d{\bf x}\,\varphi^3({\bf x},t), \label{e66}
\end{eqnarray}
where we have integrated by parts: $\int d{\bf x}\,(\nabla_{\!\bf
x}\varphi)^2=- \int d{\bf x}\,\varphi\nabla_{\!\bf x}^2\varphi$.
We define the inverse Green's function $G_{{\bf x}{\bf y}}^{-1}$
by $G_{{\bf x}{\bf y}}^{-1}\equiv(\mu^2- \nabla_{\!\bf
x}^2)\delta({\bf x}-{\bf y})$, so that $G_{{\bf x}{\bf y}}=(\mu^2-
\nabla_{\!\bf x}^2)^{-1}\delta({\bf x}-{\bf y})$ and
$\int\!\!d{\bf z}\,G_{{\bf x}{\bf z}}^{-1}G_{{\bf z}{\bf
y}}=\delta({\bf x}-{\bf y})$. Thus, the commutator condition
(\ref{e66}) reads
\begin{eqnarray}
\left[\half\int\!\!d{\bf x}\,\pi^2({\bf x},t)+\half\int d{\bf
x}\,d{\bf y}\, \varphi({\bf x},t)G_{{\bf x}{\bf
y}}^{-1}\varphi({\bf y},t),Q_1\right]=-2i\int \!\!d{\bf
x}\,\varphi^3({\bf x},t). \label{e67}
\end{eqnarray}

Equation (\ref{e67}) states that when the operator $Q_1$ is
commuted with quadratic structures of the form $\pi^2({\bf x},t)$
and $\varphi({\bf x},t) \varphi({\bf y},t)$, it must produce the
cubic term $\varphi^3({\bf x},t)$. Furthermore, the symmetry
requirements on $Q_1$ that arise from (\ref{e2}) and (\ref{e3})
imply that $Q_1$ is an even functional of $\varphi({\bf x},t)$ and
an odd functional of $\pi({\bf x},t)$. These observations allow us
to deduce an {\it ansatz} for $Q_1$ that has the same structure as
that in (\ref{e35}):
\begin{eqnarray}
Q_1=\int\!\!\!\!\int\!\!\!\!\int\!\!d{\bf x}\,d{\bf y}\,d{\bf
z}\,M_{({\bf xyz}) }\pi_{\bf x}\pi_{\bf y}\pi_{\bf
z}+\int\!\!\!\!\int\!\!\!\!\int\!\!d{\bf x}\, d{\bf y}\,d{\bf
z}\,N_{{\bf x}({\bf yz})}\varphi_{\bf y}\pi_{\bf x}\varphi_{\bf
z}, \label{e68}
\end{eqnarray}
where we have suppressed the time variable $t$ in the fields and
for brevity have indicated spatial dependences with subscripts. In
(\ref{e68}) the unknown functions $M$ and $N$ have three arguments
each. The function $M$ is totally symmetric in its three
arguments, and to emphasize this symmetry we use the notation
$M_{({\bf x}{\bf y}{\bf z})}$; $N$ is symmetric under the
interchange of the second and third arguments, and to emphasize
this symmetry we write $N_{{\bf x}({\bf yz})}$. The functions $M$
and $N$ are like form factors because they describe the spatial
distribution of the three-point interactions of the fields in
$Q_1$. We will see that the interaction of the fields is spatially
nonlocal; this nonlocality is an intrinsic property of the
operator $\mathcal{C}$.

We now proceed to determine $M$ and $N$. We substitute the {\it
ansatz} (\ref{e68}) into the commutator (\ref{e67}) and find after
some algebra that two operator identities must hold:
\begin{eqnarray}
\int\!\!\!\!\int\!\!\!\!\int\!\!\!\!\int\!\!d{\bf x}\,d{\bf
y}\,d{\bf z}\,d{\bf w}\,N_{{\bf x}({\bf y}{\bf z})}G^{-1}_{{\bf
w}{\bf x}}\varphi_{\bf y}\varphi_{ \bf w}\varphi_{\bf
z}=-2\int\!\!d{\bf w}\,\varphi_{\bf w}^3, \label{e69}
\end{eqnarray}
\begin{eqnarray}
\int\!\!\!\!\int\!\!\!\!\int\!\!d{\bf x}\,d{\bf y}\,d{\bf
z}\,N_{{\bf x}({\bf yz})}\left(\pi_{\bf x}\pi_{\bf y}\varphi_{\bf
z}+\varphi_{\bf z}\pi_{\bf x}\pi_{ \bf
y}\right)=3\int\!\!\!\!\int\!\!\!\!\int\!\!\!\!\int\!\!d{\bf
x}\,d{\bf y}\, d{\bf z}\,d{\bf w}\,M_{({\bf xyz})}G^{-1}_{{\bf
xw}}\pi_{\bf y}\varphi_{\bf w} \pi_{\bf z}. \label{e70}
\end{eqnarray}
By commuting (\ref{e69}) three times with $\pi$, and (\ref{e70})
once with $\pi$ and twice with $\varphi$, we translate these two
operator identities into two coupled partial differential
equations for $M$ and $N$:
\begin{eqnarray}
(\mu^2-\nabla_{\!\bf x}^2)N_{{\bf x}({\bf
yz})}+(\mu^2-\nabla_{\!\bf y}^2) N_{{\bf y}({\bf
xz})}+(\mu^2-\nabla_{\!\bf z}^2)N_{{\bf z}({\bf xy})}
=-6\delta({\bf x}-{\bf y})\delta({\bf x}-{\bf z}), \label{e71}
\end{eqnarray}
\begin{eqnarray}
N_{{\bf x}({\bf y}{\bf z})}+N_{{\bf y}({\bf x}{\bf z})}=3
(\mu^2-\nabla_{\!\bf z}^2)M_{({\bf xyz})}. \label{e72}
\end{eqnarray}

To solve the system of coupled differential equations (\ref{e71})
and (\ref{e72}), we Fourier transform to momentum space, denoting
the $D$-dimensional Fourier transform of a function $f_{\bf x}$ by
$\tilde f_{\bf p} \equiv\int\!\!d{\bf x}\,f_{\bf x} e^{i{\bf
p}\cdot{\bf x}}$. Fourier transformation is effective here because
it converts the differential equations (\ref{e71}) and (\ref{e72})
into algebraic equations:
\begin{eqnarray}
\frac{1}{\tilde G_{\bf p}}\tilde N_{{\bf p}({\bf q}{\bf
r})}+\frac{1}{\tilde G_{ \bf q}}\tilde N_{{\bf q}({\bf p}{\bf
r})}+\frac{1}{\tilde G_{\bf r}}\tilde N_{ {\bf r}({\bf p}{\bf
q})}=-6(2\pi)^D\delta({\bf p}+{\bf q}+{\bf r}), \label{e73}
\end{eqnarray}
\begin{eqnarray}
\tilde N_{{\bf p}({\bf q}{\bf r})}+\tilde N_{{\bf q}({\bf p}{\bf
r})}= \frac{3}{\tilde G_{\bf r}}\tilde M_{({\bf p}{\bf q}{\bf
r})}, \label{e74}
\end{eqnarray}
where ${\tilde G_{\bf p}}=({\bf p}^2+\mu^2)^{-1}$.

Note that the right side of (\ref{e73}) contains the factor
$\delta({\bf p}+{\bf q}+{\bf r})$, which implies that the two
three-point functions $M$ and $N$ conserve momentum. We thus
introduce {\it reduced} representations of these vertex functions
in which we have factored off the delta function:
\begin{eqnarray}
\tilde M_{({\bf p}{\bf q}{\bf r})}=(2\pi)^D\tilde m_{({\bf p}{\bf
q}{\bf r})} \delta({\bf p}+{\bf q}+{\bf r}),\quad \tilde N_{{\bf
p}({\bf q}{\bf r})}= (2\pi)^D\tilde n_{{\bf p}({\bf q}{\bf
r})}\delta({\bf p}+{\bf q}+{\bf r}). \label{e75}
\end{eqnarray}
The functions $\tilde m$ and $\tilde n$ satisfy the following
algebraic equations:
\begin{eqnarray}
\frac{1}{\tilde G_{\bf p}}\tilde n_{{\bf p}({\bf
qr})}+\frac{1}{\tilde G_{\bf q} }\tilde n_{{\bf q}({\bf
pr})}+\frac{1}{\tilde G_{\bf r}}\tilde n_{{\bf r}({\bf pq})}=-6,
\label{e76}
\end{eqnarray}
\begin{eqnarray}
{\tilde G_{\bf r}}\tilde n_{{\bf p}({\bf qr})}+{\tilde G_{\bf
r}}\tilde n_{{\bf q}({\bf pr})}=3\tilde m_{({\bf pqr})}.
\label{e77}
\end{eqnarray}

There are two ways to solve these equations. A physically
transparent but longer procedure making use of tree graphs is
given in the Appendix \ref{app}. A shorter analytical approach is
presented here. We begin by noting that the right side of
(\ref{e77}) is totally symmetric in its indices. Thus, we can
obtain two new equations by permuting the indices:
\begin{eqnarray}
{\tilde G_{\bf q}}\tilde n_{{\bf p}({\bf qr})}+{\tilde G_{\bf
q}}\tilde n_{{\bf r}({\bf pq})}=3\tilde m_{({\bf pqr})},\quad
{\tilde G_{\bf p}}\tilde n_{{\bf q}( {\bf pr})}+{\tilde G_{\bf
p}}\tilde n_{{\bf r}({\bf pq})}=3\tilde m_{({\bf pqr}) }.
\label{e78}
\end{eqnarray}
We now have a sufficient number of algebraic equations, namely
(\ref{e76} -- \ref{e78}), to solve for $\tilde m$ and $\tilde n$.
The final results for $M$ and $N$ are
\begin{eqnarray}
\tilde M_{({\bf pqr})}=\frac{4\tilde G_{\bf p}^2\tilde G_{\bf
q}^2\tilde G_{\bf r}^2}{\tilde G_{\bf p}^2\tilde G_{\bf
q}^2+\tilde G_{\bf p}^2\tilde G_{\bf r}^2+ \tilde G_{\bf
q}^2\tilde G_{\bf r}^2-2\tilde G_{\bf p}\tilde G_{\bf q}\tilde G_{
\bf r}(\tilde G_{\bf p}+\tilde G_{\bf q}+\tilde G_{\bf
r})}(2\pi)^D\delta({\bf p }+{\bf q}+{\bf r}), \label{e79}
\end{eqnarray}
\begin{eqnarray}
\tilde N_{{\bf p}({\bf qr})}=\frac{6\tilde G_{\bf p}\tilde G_{\bf
q}\tilde G_{ \bf r}(\tilde G_{\bf p}\tilde G_{\bf r}+\tilde G_{\bf
p}\tilde G_{\bf q}-\tilde G_{\bf q}\tilde G_{\bf r})}{\tilde
G_{\bf p}^2\tilde G_{\bf q}^2+\tilde G_{\bf p}^2\tilde G_{\bf
r}^2+\tilde G_{\bf q}^2\tilde G_{\bf r}^2-2\tilde G_{\bf p} \tilde
G_{\bf q}\tilde G_{\bf r}(\tilde G_{\bf p}+\tilde G_{\bf q}+\tilde
G_{\bf r})}(2\pi)^D\delta({\bf p}+{\bf q}+{\bf r}). \label{e80}
\end{eqnarray}
As a check of these results we compare (\ref{e79}) and (\ref{e80})
with (\ref{e36}), which describes the case $D=0$ (quantum
mechanics). When $D=0$, we have just $\tilde G_{\bf p}=\mu^{-2}$.
Substituting this expression for $\tilde G$ into (\ref{e79}) and
(\ref{e80}), we find that these equations reduce exactly to
(\ref{e36}).

Next, we substitute $\tilde G_{\bf p}=({\bf p}^2+\mu^2)^{-1}$ into
(\ref{e79}) and (\ref{e80}) and use the inverse Fourier transform
$f_{\bf x}=(2\pi)^{-D} \int d{\bf p}\,\tilde f_{\bf p} e^{-i{\bf
p}\cdot {\bf x}}$ to express $M$ and $N$ in coordinate space:
\begin{eqnarray}
M_{({\bf xyz})}=\int\!\!\!\!\int\!\!\!\!\int\frac{d{\bf p}\,d{\bf
q}\,d{\bf r}} {(2\pi)^{3D}}\,\frac{4 e^{-i{\bf x}\cdot{\bf
p}-i{\bf y}\cdot{\bf q}-i{\bf z} \cdot{\bf r}}(2\pi)^D\delta({\bf
p}+{\bf q}+{\bf r})}{\mathcal{D}({\bf p},{\bf q},{\bf r})},
\label{e81}
\end{eqnarray}
\begin{eqnarray}
N_{{\bf x}({\bf yz})}=\int\!\!\!\!\int\!\!\!\!\int\frac{d{\bf
p}\,d{\bf q}\,d {\bf r}}{(2\pi)^{3D}}\,\frac{6e^{-i{\bf
x}\cdot{\bf p}-i{\bf y}\cdot{\bf q}-i{ \bf z}\cdot{\bf
r}}(2\pi)^D\delta({\bf p}+{\bf q}+{\bf r})({\bf q}^2+{\bf r}^2-{
\bf p}^2+\mu^2)}{\mathcal{D}({\bf p},{\bf q},{\bf r})}.
\label{e82}
\end{eqnarray}
where $\mathcal{D}({\bf p},{\bf q},{\bf r})={\bf p}^4+{\bf
q}^4+{\bf r}^4 -2({\bf p}^2{\bf q}^2+{\bf q}^2{\bf r}^2+{\bf
r}^2{\bf p}^2)-2\mu^2({\bf p}^2+ {\bf q}^2+{\bf r}^2)-3\mu^4$. We
perform the ${\bf r}$ integral in (\ref{e81}) and (\ref{e82})
using the delta function and obtain
\begin{eqnarray}
M_{({\bf xyz})}=-\frac{4}{(2\pi)^{2D}}\!\int\!\!\!\!\int\!\!d{\bf
p}\,d{\bf q} \,\frac{e^{i({\bf x}-{\bf y})\cdot{\bf p}+i({\bf
x}-{\bf z})\cdot{\bf q}}} {{\mathcal{D}}({\bf p},{\bf q})},
\label{e83}
\end{eqnarray}
where ${\mathcal{D}}({\bf p},{\bf q})=4[{\bf p}^2{\bf q}^2-({\bf
p}\cdot{\bf q}) ^2]+4\mu^2({\bf p}^2+{\bf p}\cdot{\bf q}+{\bf
q}^2)+3\mu^4$ is positive, and
\begin{eqnarray}
N_{{\bf x}({\bf yz})}=3\left(\nabla_{\!\bf y}\cdot\nabla_{\!\bf
z}+\half\mu^2 \right)M_{({\bf xyz})}. \label{e84}
\end{eqnarray}

\subsection{The $(1+1)$-Dimensional Case}

For general $D$ it is difficult to evaluate the double integral
(\ref{e83}) in closed form. However, when $D=1$
[$(1+1)$-dimensional quantum field theory] we can evaluate the
integral because the quartic terms in ${\mathcal{D}}({\bf p}, {\bf
q})$ cancel. The evaluation procedure exploits the strict
positivity of the denominator $\mathcal{D}(p,q)$,
$${\mathcal{D}}(p,q)=4\mu^2(p^2+pq+q^2)+3\mu^4=2\mu^2
[p^2+q^2+(p+q)^2]+3\mu^4>0, $$ to construct the one-dimensional
integral identity $\mathcal{D}^{-1}=\int_0 ^\infty
dt\,e^{-\mathcal{D}t}$ $(\mathcal{D}>0)$. This identity allows us
to rewrite $M_{(xyz)}$ as the triple integral
\begin{eqnarray}
M_{(xyz)}=-\frac{4}{\mu^2(2\pi)^2}\int\!\!\!\!\int\!\!dp\,dq
\int_{t=0}^\infty dt\,e^{-3\mu^2
t}e^{i(x-y)p+i(x-z)q-4t(p^2+pq+q^2)}. \label{e85}
\end{eqnarray}
Note that when $D=1$ the variables $x$, $y$, $z$, $p$, $q$, and so
on, are scalars and not vectors, so we no longer use boldface
notation.

To evaluate this integral we first complete the square in the $q$
variable in the exponent and translate the $q$ integration
variable by $q\to q-\frac{p}{2}+ \frac{i(x-z)}{8t}$. We then
complete the square in the $p$ integration variable and translate
$p$ by $p\to p-\frac{i(x-z)}{12t}+\frac{i(x-y)}{6t}$. This gives
\begin{eqnarray}
M_{(xyz)}=-\frac{4}{\mu^2(2\pi)^2}\int\!\!\!\!\int\!\!dp\,dq
\int_{t=0}^\infty dt\,e^{-3\mu^2 t -3tp^2-4tq^2-\rho^2/(12t)},
\label{e86}
\end{eqnarray}
where $\rho$, which is totally symmetric in $x$, $y$, and $z$, is
the positive square root of
\begin{eqnarray}
\rho^2=\half[(x-y)^2+(y-z)^2+(z-x)^2]. \label{e87}
\end{eqnarray}

We now perform the scalings $p\to p/\sqrt{3t}$ and $q\to
q/\sqrt{4t}$. The result is that the integral (\ref{e86})
representing $M_{(xyz)}$ factors into three one-dimensional
integrals:
\begin{eqnarray}
M_{(xyz)}=-4\mu^{-2}(2\pi)^{-2}12^{-1/2}I^2J, \label{e88}
\end{eqnarray}
where $I$ is the Gaussian integral $I=\int
dq\,e^{-q^2}=\sqrt{\pi}$ and $J=\int_{t=0}^\infty
dt\,t^{-1}e^{-3\mu^2t-\rho^2/( 12t)}$. Finally, we use the
integral representation \cite{GRAD}
$$\int_{t=0}^\infty dt\,e^{-t-a^2/t}t^{-1}=2{\rm K}_0(2a),$$
where ${\rm K}_0$ is the associated Bessel function. Thus,
$J=2{\rm K}_0(\mu \rho)$. Combining the factors in (\ref{e88}), we
find that for a $(1+1)$-dimensional quantum field theory
(\ref{e83}) evaluates to
\begin{eqnarray}
M_{(xyz)}=-\frac{1}{\pi\sqrt{3}\mu^2}{\rm K}_0(\mu\rho).
\label{e89}
\end{eqnarray}

Next we calculate $N$ using (\ref{e84}). The result is
\begin{eqnarray}
N_{x(yz)}&=&-\frac{3\sqrt{3}}{4\pi}\left[1-\frac{(y-z)^2}
{\rho^2}\right]{\rm K}_
0(\mu\rho)+\frac{\sqrt{3}}{\pi}\left[1-\frac{3(y-z)^2}
{2\rho^2}\right]\frac{ {\rm K}_0^\prime(\mu\rho)}{\mu\rho}
\nonumber\\ && \nonumber\\
&&\quad+\frac{1}{\mu^2}\left[1-\frac{3(y-z)^2}
{\rho^2}\right]\delta(x-y)\delta(x -z). \label{e90}
\end{eqnarray}

The mathematics underlying the solutions in (\ref{e89}) and
(\ref{e90}) is rather subtle and bears further discussion. First,
it is important to mention that while we have expressed $M$ and
$N$ as functions of the three variables $x$, $y$, and $z$,
translation invariance implies that these functions really depend
on only two variables, say, the differences $x-y$ and $x-z$. We
therefore define the two variables $\eta$ and $\zeta$ by
\begin{eqnarray}
\eta=x-\half(y+z)\quad{\rm
and}\quad\zeta={\textstyle\frac{\sqrt{3}}{2}}(y-z). \label{e91}
\end{eqnarray}
In terms of these new variables we have
\begin{eqnarray}
\rho^2&=&\eta^2+\zeta^2,\nonumber\\
\partial_x^2+\partial_y^2+\partial_z^2&=&
{\textstyle\frac{3}{2}}(\partial_\eta^2
+\partial_\zeta^2)={\textstyle\frac{3}{2}}
\nabla_{\eta,\zeta}^2,\nonumber\\
{\textstyle\frac{2}{\sqrt{3}}}\delta(x-y)\delta(x-z)&=&
\delta(\eta)\delta(\zeta)
={\textstyle\frac{1}{2\pi\rho}}\delta(\rho). \label{e92}
\end{eqnarray}

The reason for introducing new variables and for emphasizing that
we are working in the two-dimensional $(\eta,\zeta)$ space is that
in {\it two-dimensional} space the associated Bessel function
$\frac{1}{2\pi}{\rm K}_0(\mu\rho)$ is the Green's function:
\begin{eqnarray}
\left(\mu^2-\nabla_{\eta,\zeta}^2\right)
{\textstyle\frac{1}{2\pi}}{\rm K_0}(\mu
\rho)=\delta(\eta)\delta(\zeta)={\textstyle\frac{1}
{2\pi\rho}}\delta(\rho). \label{e93}
\end{eqnarray}
This equation explains the appearance of the contact term
(delta-function term) in the expression for $N$ in (\ref{e90}).
The delta function is rotationally symmetric because the Green's
function is rotationally symmetric and thus we can replace
$\nabla_{\eta,\zeta}^2$ in (\ref{e93}) by $d^2/d\rho^2+d/d\rho$.
Hence, we see that two derivatives of ${\rm K}_0(\mu\rho)$ give
rise to a delta function:
\begin{eqnarray}
{\rm K}_0^{\prime\prime}(\mu\rho)={\rm
K}_0(\mu\rho)-{\textstyle\frac{1}{\mu\rho }}{\rm
K}_0^\prime(\mu\rho)-{\textstyle\frac{1}{\mu^2\rho}} \delta(\rho).
\label{e94}
\end{eqnarray}

We have checked that $M$ and $N$ in (\ref{e89}) and (\ref{e90})
satisfy the partial differential equations (\ref{e71}) and
(\ref{e72}). We can verify (\ref{e72}) by direct differentiation.
To verify (\ref{e71}) we use the variables $\eta$ and $\zeta$ and
take the indicated derivatives. The result is a combination of
${\rm K}_0(\mu\rho)$, ${\rm K}_0^\prime(\mu\rho)$, $\delta(\rho
)$, $\delta^\prime(\rho)$, and $\delta^{\prime\prime}(\rho)$
terms. The coefficients of all of these terms vanish except for
the coefficient $\delta( \rho)$ term, and this coefficient
reproduces exactly the right side of (\ref{e71}).

Because our formulas for $M$ and $N$ involve Bessel functions, we
see clearly that $Q_1$ represents a {\it nonlocal} interaction of
three fields. However, as associated Bessel functions decrease
exponentially rapidly for large argument, the degree of
nonlocality is small.

\section{Quantum Field Theory with Several Interacting Fields}
\label{s8}

The field theoretic calculations in Sec.~\ref{s7} can be extended
to cubic quantum field theories having two and three interacting
scalar fields.

\subsection{$\varphi_1\varphi_2^2$ Theory}
\label{ssss1}

We consider first the case of two scalar fields $\varphi_{\bf
x}^{(1)}$ and $\varphi_{\bf x}^{(2)}$ whose dynamics is described
by the Hamiltonian $H=H_0^{(1)}+H_0^{(2)}+\epsilon H_1$, where
\begin{eqnarray}
H_0^{(j)}=\half\int\!\!d{\bf x}\,\left(\pi_{\bf x}^{(j)}\right)^2
+\half\int\!\!\!\!\int\!\!d{\bf x}\,d{\bf y}\,\left(G_{\bf
xy}^{(j)} \right)^{-1} \varphi_{\bf x}^{(j)}\varphi_{\bf
y}^{(j)}\qquad(j=1,\,2) \label{e95}
\end{eqnarray}
and $H_1=i\int\!\!d{\bf x}\,\big(\varphi_{\bf
x}^{(1)}\big)^2\varphi_{\bf x}^ {(2)}$. The Green's function
$G_{{\bf x}{\bf y}}^{(j)}$ is the solution to the equation
$$\left(\mu_j^2-\nabla_{\!\bf x}^2\right)G_{\bf xy}^{(j)}=
\delta({\bf x}-{\bf y})\quad(j=1,2).$$ This quantum field theory
is the analog of the quantum mechanical theory in (\ref{e11}).

To determine $\mathcal{C}$ to order $\epsilon$ we need to solve
the operator equation
\begin{eqnarray}
\left[H_0^{(1)}+H_0^{(2)},Q_1\right]=-2H_1, \label{e96}
\end{eqnarray}
which is the two-field generalization of (\ref{e66}). To find the
solution to this equation we make an {\it ansatz} analogous to
that in (\ref{e41}) for the operator $Q_1$:
$$Q_1=\int\!\!\!\!\int\!\!\!\!\int\!\!d{\bf x}\,d{\bf y} \,d{\bf
z}\big[M_{{\bf x}({\bf yz})}\pi_{\bf x}^{(2)}\pi_{\bf
y}^{(1)}\pi_{\bf z}^{(1)}+N_{\bf xyz}^{ (1)}\!\left(\pi_{\bf
z}^{(1)}\varphi_{\bf y}^{(1)}\!+\!\varphi_{\bf y}^{(1)} \pi_{\bf
z}^{(1)}\right)\!\varphi_{\bf x}^{(2)}+N_{{\bf x}({\bf yz})
}^{(2)}\pi_{\bf x}^{(2)}\varphi_{\bf y}^{(1)}\varphi_{\bf
z}^{(1)}\big],$$ where $M_{{\bf x}({\bf yz})}$, $N_{\bf
xyz}^{(1)}$, and $N_{{\bf x}({\bf yz})}^{ (2)}$ are unknown
functions of three arguments each. As indicated by the
parentheses, $M$ and $N^{(2)}$ are symmetric in their last two
arguments.

We then substitute $Q_1$ into (\ref{e96}) and use the following
identities to perform the algebra:
\begin{eqnarray}
\left[H_0^{(1)},Q_1\right]&=&-i\int\!\!\!\!\int\!\!\!\!\int\!\!
d{\bf x}\,d{\bf y}\,d{\bf z}\big[2N_{\bf xyz}^{(1)}\pi_{\bf
y}^{(1)}\pi_{\bf z}^{(1)}\varphi_{ \bf x}^{(2)}+N_{{\bf x}({\bf
y}{\bf z})}^{(2)}\pi_{\bf x}^{(2)}\left(\varphi_{ \bf
y}^{(1)}\pi_{\bf z}^{(1)}+\pi_{\bf z}^{(1)}\varphi_{\bf
y}^{(1)}\right)\big]
\nonumber\\
&&\quad +i\int\!\!\!\!\int\!\!\!\!\int\!\!\!\!\int\!\!d{\bf
w}\,d{\bf x}\,d{\bf y}\,d{\bf z}\left(G_{{\bf w}{\bf
y}}^{(1)}\right)^{-1}M_{{\bf x}({\bf y}{\bf z}) }\pi_{\bf
x}^{(2)}\left(\varphi_{\bf w}^{(1)}\pi_{\bf z}^{(1)}+
\pi_{\bf z}^{(1)}\varphi_{\bf w}^{(1)}\right)\nonumber\\
&&\quad+2i\int\!\!\!\!\int\!\!\!\!\int\!\!\!\!\int\!\!d{\bf
w}\,d{\bf x}\,d{\bf y}\,d{\bf z}\left(G_{{\bf w}{\bf
z}}^{(1)}\right)^{-1}N_{\bf xyz}^{(1)}\varphi_{ \bf
x}^{(2)}\varphi_{\bf w}^{(1)}\varphi_{\bf y}^{(1)},\nonumber\\
\left[H_0^{(2)},Q_1\right]&=&-i\int\!\!\!\!\int\!\!\!\!\int\!\!
d{\bf x}\,d{\bf y}\,d{\bf z}N_{\bf xyz}^{(1)}\left(\pi_{\bf
z}^{(1)}\varphi_{\bf y}^{(1)}+ \varphi_{\bf y}^{(1)}\pi_{\bf
z}^{(1)}\right)\pi_{\bf x}^{(2)} \nonumber\\
&&\quad+i\int\!\!\!\!\int\!\!\!\!\int\!\!\!\!\int\!\!d{\bf
w}\,d{\bf x}\,d{\bf y}\,d{\bf z}\left(G_{{\bf w}{\bf
x}}^{(2)}\right)^{-1}N_{{\bf x}({\bf y}{\bf z})
}^{(2)}\varphi_{\bf y}^{(1)}\varphi_{\bf z}^{(1)}\varphi_{\bf
w}^{(2)}
\nonumber\\
&&\quad+i\int\!\!\!\!\int\!\!\!\!\int\!\!\!\!\int\!\!d{\bf
w}\,d{\bf x}\,d{\bf y }\,d{\bf z}\left(G_{{\bf w}{\bf
x}}^{(2)}\right)^{-1}M_{{\bf x}({\bf y}{\bf z})} \pi_{\bf
y}^{(1)}\pi_{\bf z}^{(1)}\varphi_{\bf w}^{(2)}. \label{e97}
\end{eqnarray}
Substituting (\ref{e97}) into (\ref{e96}) gives an operator
equation involving the unknown functions $M$, $N^{(1)}$, and
$N^{(2)}$. We then convert these operator equations into a system
of three coupled partial differential equations by commuting with
products of three fields:
\begin{eqnarray}
\left(\mu_1^2-\nabla_{\!\bf z}^2\right)\!N_{\bf
xyz}^{(1)}+\left(\mu_1^2-\nabla_ {\!\bf y}^2\right)\!N_{\bf
xyz}^{(1)}+\left(\mu_2^2-\nabla_{\!\bf x}^2\right)\! N_{{\bf
x}({\bf yz})}^{(2)}&=&-2\delta({\bf x}-{\bf y})
\delta({\bf x}-{\bf z}),\nonumber\\
\left(\mu_1^2-\nabla_{\!\bf y}^2\right)\!M_{{\bf x}({\bf
yz})}&=&N_{\bf xyz}^{(1
)}+N_{{\bf x}({\bf yz})}^{(2)},\nonumber\\
\left(\mu_2^2-\nabla_{\!\bf x}^2\right)\!M_{{\bf x}({\bf
yz})}&=&N_{\bf xyz}^{(1 )}+N_{\bf xzy}^{(1)}. \label{e98}
\end{eqnarray}

To solve this system of partial differential equations we perform
Fourier transforms in the variables ${\bf x}$, ${\bf y}$, and
${\bf z}$ and obtain
\begin{eqnarray}
\frac{1}{\tilde G_{\bf r}^{(1)}}\tilde N_{\bf
pqr}^{(1)}+\frac{1}{\tilde G_{\bf q}^{(1)}}\tilde N_{\bf
prq}^{(1)}+\frac{1}{\tilde G_{\bf p}^{(2)}}\tilde N_{{\bf p}({\bf
qr})}^{(2)}&=&-2(2\pi)^D\delta({\bf p}+{\bf q}+{\bf r}),
\nonumber\\ \frac{1}{\tilde G_{\bf q}^{(1)}}\tilde M_{{\bf p}({\bf
qr})}&=&\tilde N_{\bf pqr
}^{(1)}+\tilde N_{{\bf p}({\bf qr})}^{(2)},\nonumber\\
\frac{1}{\tilde G_{\bf p}^{(2)}}\tilde M_{{\bf p}({\bf qr})}
&=&\tilde N_{\bf pqr}^{(1)}+\tilde N_{\bf prq}^{(1)}. \label{e99}
\end{eqnarray}
Note that in (\ref{e99}) we have three linear algebraic equations
in the four unknowns $M_{{\bf p}({\bf qr})}$, $N_{\bf pqr}^{(1)}$,
$N_{\bf prq}^{(1)}$, and $N_{{\bf p}({\bf qr})}^{(2)}$. Thus, we
construct another equation from the second of the three equations
in (\ref{e99}) by interchanging the momenta ${\bf q}$ and ${\bf
r}$:
\begin{eqnarray}
\frac{1}{\tilde G_{\bf r}^{(1)}}\tilde M_{{\bf p}({\bf
qr})}=\tilde N_{\bf prq}^{(1)}+\tilde N_{{\bf p}({\bf qr})}^{(2)}.
\label{e100}
\end{eqnarray}

Solving the algebraic equations (\ref{e99}) and (\ref{e100}), we
obtain the following expressions for $\tilde M$, $\tilde N^{(1)}$,
and $\tilde N^{(2)}$:
\begin{eqnarray}
\tilde M_{{\bf p}({\bf qr})}&=&\frac{4}{\mathcal{D}({\bf p},{\bf
q},{\bf r})}
(2\pi)^D\delta({\bf p}+{\bf q}+{\bf r}),\nonumber\\
\tilde N_{\bf pqr}^{(1)}&=&\frac{2\big({\bf p}^2+{\bf q}^2-{\bf
r}^2+\mu_2^2 \big)}{\mathcal{D}({\bf p},{\bf q},{\bf
r})}(2\pi)^D\delta({\bf p}+{\bf q}+
{\bf r}),\nonumber\\
\tilde N_{{\bf p}({\bf qr})}^{(2)}&=&\frac{2\big[{\bf q}^2+{\bf
r}^2-{\bf p}^2+2 \mu_1^2-\mu_2^2\big]}{\mathcal{D}({\bf p},{\bf
q},{\bf r})}(2\pi)^D\delta({\bf p}+{\bf q}+{\bf r}), \label{e101}
\end{eqnarray}
where the denominator $\mathcal{D}({\bf p},{\bf q},{\bf r})$ is
given by
$$\mathcal{D}({\bf p},{\bf q},{\bf r})={\bf p}^4+{\bf q}^4+ {\bf
r}^4-2\big({\bf p}^2{\bf q}^2+{\bf p}^2{\bf r}^2+{\bf q}^2{\bf
r}^2\big)+2\mu_2^2\left({\bf p}^2 -{\bf q}^2-{\bf
r}^2\right)-4\mu_1^2{\bf p}^2+\mu_2^4-4\mu_1^2\mu_2^2.$$ Observe
that if we set $\mu_1=\mu_2=1$ and take the quantum mechanical
limit $D\to0$, Eq.~(\ref{e101}) reduces to (\ref{e42}).

Finally, we transform (\ref{e101}) back to coordinate space by
calculating the inverse Fourier transforms. The integrals to be
performed are triple $D$-dimensional integrals, but we can perform
the integral over ${\bf p}$ by using the delta function. We get
\begin{eqnarray}
M_{{\bf x}({\bf
yz})}=-\frac{4}{(2\pi)^{2D}}\!\int\!\!\!\!\int\!\!d{\bf q}\, d
{\bf r}\,\frac{e^{i({\bf x}-{\bf y})\cdot{\bf q}+i({\bf x}-{\bf
z})\cdot{\bf r} }}{{\mathcal{D}}({\bf q},{\bf r})}, \label{e102}
\end{eqnarray}
where ${\mathcal{D}}({\bf q},{\bf r})=4[{\bf q}^2{\bf r}^2-({\bf
q}\cdot{\bf r}) ^2]+4\mu_1^2({\bf q}+{\bf r})^2-4\mu_2^2{\bf
q}\cdot{\bf r}-\mu_2^4+4\mu_1^2 \mu_2^2$, and
\begin{eqnarray}
N_{\bf xyz}^{(1)}&=&\left[-\nabla_{\!\bf y}^2-\nabla_{\!\bf
y}\cdot\nabla_{\!\bf
z}+\half\mu_2^2\right]M_{{\bf x}({\bf yz})},\nonumber\\
N_{{\bf x}(\bf{yz})}^{(2)}&=&\left[\nabla_{\!\bf
y}\cdot\nabla_{\!\bf z}- \mu_1^2-\half\mu_2^2\right]M_{ {\bf
x}({\bf yz})}. \label{e103}
\end{eqnarray}

We mention that for the special case $D=1$ [quantum field theory
in ($1+1$)-dimensional space-time], the quartic terms in the
denominator ${\mathcal{D}}({\bf q},{\bf r})$ again drop out and it
is possible to evaluate the integral in (\ref{e102}) in terms of
Bessel functions using the the integration techniques described in
Sec.~\ref{s7}. For this special case we obtain
\begin{eqnarray}
M_{x(yz)}=-\frac{1}{\pi\mu_2\sqrt{4\mu_1^2-\mu_2^2}}{\rm
K}_0(\mu_2\rho), \label{e104}
\end{eqnarray}
where
$4\mu_2^2\rho^2=\mu_2^2(2x-y-z)^2+(4\mu_1^2-\mu_2^2)(y-z)^2$. The
result in (\ref{e104}) reduces to that in (\ref{e89}) in the
equal-mass case $\mu_1= \mu_2=\mu$. Also, our results in $D=1$ for
$N_{xyz}^{(1 )}$ and $N_{x(yz)}^{( 2)}$ are
\begin{eqnarray}
N_{xyz}^{(1)}&=&-\frac{\sqrt{4\mu_1^2-\mu_2^2}}{4\pi\mu_2}
\frac{(x-z)(y-z)}{\rho
^2}{\rm K}_0(\mu_2\rho) \nonumber\\
&&\!\!\!\!\!\!\!\!\!\!\!\!-\frac{1}{2\pi\rho\mu_2^2
\sqrt{4\mu_1^2-\mu_2^2}}\left
[\mu_2^2-(4\mu_1^2-\mu_2^2)\frac{(x-z)(y-z)}{\rho^2}\right] {\rm
K}_0^\prime( \mu_2\rho)\nonumber\\
&&\!\!\!\!\!\!\!\!\!\!\!\!-\frac{1}{\mu_2^2(4\mu_1^2-\mu_2^2)}
\left[2\mu_2^2-(4
\mu_1^2-\mu_2^2)\frac{(x-z)(y-z)}{\rho^2}\right]\delta(x-y)
\delta(x-z),
\nonumber\\ \nonumber\\
N_{x(yz)}^{(2)}&=&\frac{\sqrt{4\mu_1^2-\mu_2^2}}{4\pi\mu_2}
\frac{(x-z)(y-z)}
{\rho^2}{\rm K}_0(\mu_2\rho) \nonumber\\
&&\!\!\!\!\!\!\!\!\!\!\!\!-\frac{1}{2\pi\rho\mu_2^2
\sqrt{4\mu_1^2-\mu_2^2}}\left
[2\mu_1^2-\mu_2^2-(4\mu_1^2-\mu_2^2)\frac{(x-z)(y-z)}{\rho^2}
\right]{\rm K}_0^
\prime(\mu_2\rho)\nonumber\\
&&\!\!\!\!\!\!\!\!\!\!\!\!-\frac{1}{\mu_2^2(4\mu_1^2-\mu_2^2)}
\left[4\mu_1^2-2
\mu_2^2-(4\mu_1^2-\mu_2^2)\frac{(x-z)(y-z)}{\rho^2}
\right]\delta(x-y) \delta(x-z). \label{e105}
\end{eqnarray}

\subsection{$\varphi_1\varphi_2\varphi_3$ Theory}
\label{ssss2}

We now consider the case of {\it three} interacting scalar fields
whose dynamics is described by the Hamiltonian
\begin{eqnarray}
H=H_0^{(1)}+H_0^{(2)}+H_0^{(3)}+\epsilon H_1, \label{e106}
\end{eqnarray}
where $H_0^{(j)}$ is given in (\ref{e95}) and $H_1=i\int\!\!d{\bf
x}\,\varphi_{ \bf x}^{(1)}\varphi_{\bf x}^{(2)}\varphi_{\bf
x}^{(3)}$. This quantum field theory has the interesting property
that its perturbative solution is finite for $D<3$; there are no
divergent graphs in less than three space-time dimensions.

To find the operator $\mathcal{C}$ to leading order in $\epsilon$,
we need to solve the operator equation
\begin{eqnarray}
\big[H_0^{(1)}+H_0^{(2)}+H_0^{(3)},Q_1\big]=-2H_1. \label{e107}
\end{eqnarray}
We introduce the {\it ansatz}
\begin{eqnarray}
Q_1&=&\int\!\!\!\!\int\!\!\!\!\int\!\!d{\bf x}\,d{\bf y}\,d{\bf
z}\,N_{{\bf x}{ \bf y}{\bf z}}^{(1)}\pi_{\bf x}^{(1)}\varphi_{\bf
y}^{(2)}\varphi_{\bf z}^{(3)}+
\int\!\!\!\!\int\!\!\!\!\int\!\!d{\bf x}\,d{\bf y}\,d{\bf
z}\,N_{\bf xyz}^{(2)} \pi_{\bf x}^{(2)}\varphi_{\bf
y}^{(3)}\varphi_{\bf z}^{(1)} \nonumber\\
&&\quad+\int\!\!\!\!\int\!\!\!\!\int\!\!d{\bf x}\,d{\bf y}\,d{\bf
z}\,N_{\bf xyz }^{(3)}\pi_{\bf x}^{(3)}\varphi_{\bf
y}^{(1)}\varphi_{\bf z}^{(2)}+\int\!\!\!\!
\int\!\!\!\!\int\!\!d{\bf x}\,d{\bf y}\,d{\bf z}\,M_{\bf
xyz}\pi_{\bf x}^{(1)} \pi_{\bf y}^{(2)}\pi_{\bf z}^{(3)}.
\label{e108}
\end{eqnarray}

We then establish the following three results:
\begin{eqnarray}
\left[H_0^{(1)},Q_1\right]&=&-i\int\!\!\!\!\int\!\!\!\!\int\!\!d
{\bf x}\,d{\bf y }\,d{\bf z}\left(N_{\bf xyz}^{(2)}\pi_{\bf
z}^{(1)}\pi_{\bf x}^{(2)}\varphi_{\bf y}^{(3)}+N_{\bf
xyz}^{(3)}\pi_{\bf y}^{(1)}\pi_{\bf x}^{(3)}\varphi_{\bf z}^{(2)
}\right)\nonumber\\
&&\quad+i\int\!\!\!\!\int\!\!\!\!\int\!\!\!\!\int\!\!d{\bf
w}\,d{\bf x}\,d{\bf y }\,d{\bf z}\left(G_{{\bf w}{\bf
x}}^{(1)}\right)^{-1}N_{\bf xyz}^{(1)} \varphi_{\bf
w}^{(1)}\varphi_{\bf y}^{(2)}\varphi_{\bf z}^{(3)} \nonumber\\
&&\quad+i\int\!\!\!\!\int\!\!\!\!\int\!\!\!\!\int\!\!d{\bf
w}\,d{\bf x}\,d{\bf y}\,d{\bf z}\left(G_{{\bf w}{\bf
x}}^{(1)}\right)^{-1}M_{\bf xyz} \pi_{\bf y}^{(2)}\pi_{\bf
z}^{(3)}\varphi_{\bf w}^{(1)}, \nonumber\\
\left[H_0^{(2)},Q_1\right]&=&
-i\int\!\!\!\!\int\!\!\!\!\int\!\!d{\bf x}\,d{\bf y}\,d{\bf
z}\left(N_{\bf xyz}^{(1)}\pi_{\bf y}^{(2)}\pi_{\bf
x}^{(1)}\varphi_{ \bf z}^{(3)}+N_{\bf xyz}^{(3)}\pi_{\bf
x}^{(3)}\pi_{\bf z}^{(2)}
\varphi_{\bf y}^{(1)}\right)\nonumber\\
&&\quad+i\int\!\!\!\!\int\!\!\!\!\int\!\!\!\!\int\!\!d{\bf
w}\,d{\bf x}\,d{\bf y }\,d{\bf z}\left(G_{{\bf w}{\bf
x}}^{(2)}\right)^{-1}N_{\bf xyz}^{(2)} \varphi_{\bf
w}^{(2)}\varphi_{\bf y}^{(3)}\varphi_{\bf z}^{(1)}
\nonumber\\
&&\quad+i\int\!\!\!\!\int\!\!\!\!\int\!\!\!\!\int\!\!d{\bf
w}\,d{\bf x}\,d{\bf y }\,d{\bf z}\left(G_{{\bf w}{\bf
y}}^{(2)}\right)^{-1}M_{\bf xyz} \pi_{\bf x}^{(1)}\pi_{\bf
z}^{(3)}\varphi_{\bf w}^{(2)}, \nonumber\\
\left[H_0^{(3)},Q_1\right]&=&
-i\int\!\!\!\!\int\!\!\!\!\int\!\!d{\bf x}\,d{\bf y}\,d{\bf
z}\left(N_{\bf xyz}^{(1)}\pi_{\bf x}^{(1)}\pi_{\bf
z}^{(3)}\varphi_{ \bf y}^{(2)}+N_{\bf xyz}^{(2)}\pi_{\bf
x}^{(2)}\pi_{\bf y}^{(3)}\varphi_{\bf z}^
{(1)}\right)\nonumber\\
&&\quad+i\int\!\!\!\!\int\!\!\!\!\int\!\!\!\!\int\!\!d{\bf
w}\,d{\bf x}\,d{\bf y }\,d{\bf z}\left(G_{{\bf w}{\bf
x}}^{(3)}\right)^{-1}N_{{\bf x}{\bf y}{\bf z}}^{ (3)}\varphi_{\bf
y}^{(1)}\varphi_{\bf z}^{(2)} \varphi_{\bf w}^{(3)}\nonumber\\
&&\quad+i\int\!\!\!\!\int\!\!\!\!\int\!\!\!\!\int\!\!d{\bf
w}\,d{\bf x}\,d{\bf y }\,d{\bf z}\left(G_{{\bf w}{\bf
z}}^{(3)}\right)^{-1}M_{{\bf x}{\bf y}{\bf z}} \pi_{\bf
x}^{(1)}\pi_{\bf y}^{(2)}\varphi_{\bf w}^{(3)}. \label{e109}
\end{eqnarray}
From these equations we deduce the following system of
differential equations:
\begin{eqnarray}
(\mu_1^2-\nabla_{\!\bf x}^2)N_{\bf xyz}^{(1)}
+(\mu_2^2-\nabla_{\!\bf y}^2)N_{\bf yzx}^{(2)}
+(\mu_3^2-\nabla_{\!\bf z}^2)N_{\bf zxy}^{(3)} &=&-2\delta({\bf
x}-{\bf y})\delta({\bf x}-{\bf z}), \nonumber\\
\left(\mu_3^2-\nabla_{\!\bf z}^2\right)M_{\bf xyz}&=&N_{\bf
xyz}^{(1)}
+N_{\bf yzx}^{(2)},\nonumber\\
\left(\mu_1^2-\nabla_{\!\bf x}^2\right)M_{\bf xyz}&=&N_{\bf
yzx}^{(2)}
+N_{\bf zxy}^{(3)},\nonumber\\
\left(\mu_2^2-\nabla_{\!\bf y}^2\right)M_{\bf xyz}&=&N_{\bf
zxy}^{(3)} +N_{\bf xyz}^{(1)}. \label{e110}
\end{eqnarray}

The solutions for the unknown functions are as follows: $M_{\bf
xyz}$ is given by the integral (\ref{e83}) with the more general
formula $\mathcal{D}({\bf p}, {\bf q})=4[{\bf p}^2{\bf q}^2-({\bf
p}\cdot{\bf q})^2]+4[\mu_1^2({\bf q}^2+{\bf p}\cdot{\bf
q})+\mu_2^2({\bf p}^2+{\bf p}\cdot{\bf q})-\mu_3^2{\bf p}\cdot{\bf
q}]+m^4$ with
$3m^4=2\mu_1^2\mu_2^2+2\mu_1^2\mu_3^2+2\mu_2^2\mu_3^2-\mu_1^4-\mu_
2^4-\mu_3^4$. The $N$ coefficients are expressed as derivatives
acting on $M$:
\begin{eqnarray}
N_{\bf xyz}^{(1)}\!&=&\! \left[4\nabla_{\!\bf y}\cdot\nabla_{\!\bf
z}+2(\mu_2^2
+\mu_3^2-\mu_1^2)\right] M_{\bf xyz}, \nonumber\\
N_{\bf xyz}^{(2)}\!&=&\! \left[-4\nabla_{\!\bf
y}\cdot\nabla_{\!\bf z} -4\nabla_{\!\bf
z}^2+2(\mu_1^2+\mu_3^2-\mu_2^2)\right] M_{\bf xyz}, \nonumber\\
N_{\bf xyz}^{(3)}\!&=&\! \left[-4\nabla_{\!\bf
y}\cdot\nabla_{\!\bf z} -4\nabla_{\!\bf
y}^2+2(\mu_1^2+\mu_2^2-\mu_3^2)\right] M_{\bf xyz}. \label{e111}
\end{eqnarray}

For the case $D=1$ we have
\begin{eqnarray}
M_{(xyz)}=-\frac{1}{\pi\sqrt{3}m^2}{\rm K}_0(m\rho), \label{e112}
\end{eqnarray}
where
$2m^2\rho^2=(\mu_1^2+\mu_2^2-\mu_3^2)(x-y)^2+(\mu_2^2+\mu_3^2-\mu_1^2)
(y-z)^2+(\mu_3^2+\mu_1^2-\mu_2^2)(z-x)^2$. We also have
\begin{eqnarray}
N_{xyz}^{(1)}&=&-\frac{\sqrt{3}\,(x-y)(x-z)}{4\pi\rho^2}{\rm
K}_0(m\rho) -\frac{1}{2\pi\rho
m}\left[\frac{\mu_2^2+\mu_3^2-\mu_1^2}{\sqrt{3}\,m^2}
-\frac{\sqrt{3}\,(x-y)(x-z)}{\rho^2}\right]{\rm K}_0^\prime
(m\rho) \nonumber\\ &&\quad
-\frac{2}{m^2}\left[\frac{\mu_2^2+\mu_3^2-\mu_1^2}
{3m^2}-\frac{(x-y)(x-z)}{2\rho^2}\right]\delta(x-y)\delta(x-z),
\label{e113}
\end{eqnarray}
and analogous expressions for $N_{xyz}^{(2)}$ and $N_{xyz}^{(3)}$.

\section{Final Remarks}
\label{s9}

We have introduced a new algebraic technique for constructing the
operator ${\mathcal C}$, which is required to define the
positive-definite inner product of the Hilbert space in ${\mathcal
PT}$-symmetric quantum theories. Unlike the previously used
analytical procedure for constructing ${\mathcal C}$, which relies
on the determination of all the energy eigenstates, the algebraic
approach introduced here allows us to determine ${\mathcal C}$
directly from the operator form of the Hamiltonian. As a
consequence, the approach extends naturally to quantum field
theory. We have explicitly demonstrated the perturbative
derivation of ${\mathcal C}$ both in quantum mechanics and quantum
field theory for the case of cubic interactions.

We hope to generalize the breakthrough reported in this paper to
noncubic $\mathcal{PT}$-symmetric quantum field theories, such as
a $-g\varphi^4$ theory. A $-g\varphi^4$ quantum field theory in
four-dimensional space-time is a remarkable model because it has a
positive spectrum, is renormalizable, is asymptotically
free~\cite{BMS}, and has a nonzero one-point Green's function
$G_1=\langle\varphi\rangle$. Consequently, this theory may
ultimately be useful in elucidating the dynamics of the Higgs
sector of the standard model.

\begin{acknowledgments}
CMB is grateful to the Theoretical Physics Group at Imperial
College for its hospitality and he thanks the U.K. Engineering and
Physical Sciences Research Council, the John Simon Guggenheim
Foundation, and the U.S.~Department of Energy for financial
support. DCB gratefully acknowledges the financial support of The
Royal Society.
\end{acknowledgments}

\appendix
\section{Graphical Solution to Equations (\ref{e76}) and (\ref{e77})}
\label{app}

Graphical methods can be used to solve the simultaneous equations
(\ref{e76}) and (\ref{e77}) for $\tilde m_{({\bf pqr})}$ and
$\tilde n_{{\bf p}({\bf qr})}$. We begin by defining the unknown
function $\tilde F_{{\bf p}({\bf qr})}$ by
\begin{eqnarray}
\tilde n_{{\bf p}({\bf qr})}=\tilde G_{\bf p}\big(\tilde F_{{\bf
p}({\bf qr})} -2\big). \label{ea1}
\end{eqnarray}
In terms of $\tilde F$, (\ref{e76}) becomes
\begin{eqnarray}
\tilde F_{{\bf p}({\bf q}{\bf r})}+\tilde F_{{\bf q}({\bf p}{\bf
r})} +\tilde F_{{\bf r}({\bf p}{\bf q})}=0 \label{ea2}
\end{eqnarray}
and (\ref{e77}) becomes
\begin{eqnarray}
\tilde G_{\bf p}\tilde G_{\bf q}\tilde F_{{\bf p}({\bf q}{\bf r})}
+\tilde G_{\bf q} \tilde G_{\bf r}\tilde F_{{\bf r}({\bf p}{\bf
q})} -\tilde G_{\bf p}\tilde G_{\bf r} \tilde F_{{\bf p}({\bf
q}{\bf r})} -\tilde G_{\bf q}\tilde G_{\bf r}\tilde F_{{\bf
q}({\bf p}{\bf r})}= 2\tilde G_{\bf p}\big(\tilde G_{\bf q}-\tilde
G_{\bf r}\big), \label{ea3}
\end{eqnarray}
where we have made use of (\ref{e74}) to eliminate $\tilde m$.

We can find an exact solution to (\ref{ea2}) and (\ref{ea3}) by
using an iterative method. First, we construct a function that we
call $\tilde F_{{\bf p} ({\bf q}{\bf r})}^{(1)}$, which exactly
solves (\ref{ea2}) and reproduces the structure on the right side
of (\ref{ea3}). An {\it ansatz} that works is
\begin{eqnarray}
{\tilde F}^{(1)}_{{\bf p}({\bf q}{\bf r})} = \alpha_1\left(
\frac{{\tilde G}_{\bf q}}{{\tilde G}_{\bf r}} + \frac{{\tilde
G}_{\bf r}}{{\tilde G}_{\bf q}} \right) + \alpha_2\left(
\frac{{\tilde G}_{\bf p}}{{\tilde G}_{\bf q}} + \frac{{\tilde
G}_{\bf p}}{{\tilde G}_{\bf r}} \right) + \alpha_3\left(
\frac{{\tilde G}_{\bf r}}{{\tilde G}_{\bf p}} + \frac{{\tilde
G}_{\bf q}}{{\tilde G}_{\bf p}} \right), \label{ea4}
\end{eqnarray}
where the coefficients $\alpha_1$, $\alpha_2$, and $\alpha_3$ are
numerical parameters to be determined.

\begin{figure}[t]
\vspace{1.85in} \includegraphics{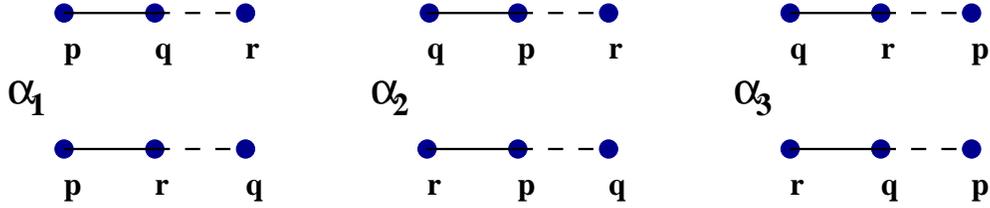} \vspace{-17mm}
\caption{Representation of $\tilde F^{(1)}_{{\bf p}({\bf q}{\bf
r})}$ in (\ref{ea4}) in terms of tree graphs. The solid lines
represent factors of $\tilde G$ and the dashed lines represent
factors of $\tilde G^{-1}$. The dots indicate points where
momentum is flowing into the graph. The momentum in the graph is
conserved and obeys the constraint ${\bf p}+{\bf q}+{\bf r}=0$.
The numerical coefficients $\alpha_1$, $\alpha_2$, and $\alpha_3$
multiply the graphs as shown in (\ref{ea4}).} \label{f2}
\end{figure}

The six terms in (\ref{ea4}) have a graphical representation (see
Fig.~\ref{f2}) in which we use a solid line to represent the
Green's function $\tilde G_{\bf p}$ and a dashed line to represent
its inverse $\tilde G_{\bf p}^ {-1}$. The dots indicate momentum
flowing into the graph. Note that the three momenta ${\bf p}$,
${\bf q}$, and ${\bf r}$ must satisfy a momentum conservation
equation ${\bf p}+{\bf q}+{\bf r}=0$.

\begin{figure}[t]
\vspace{3.65in} \includegraphics{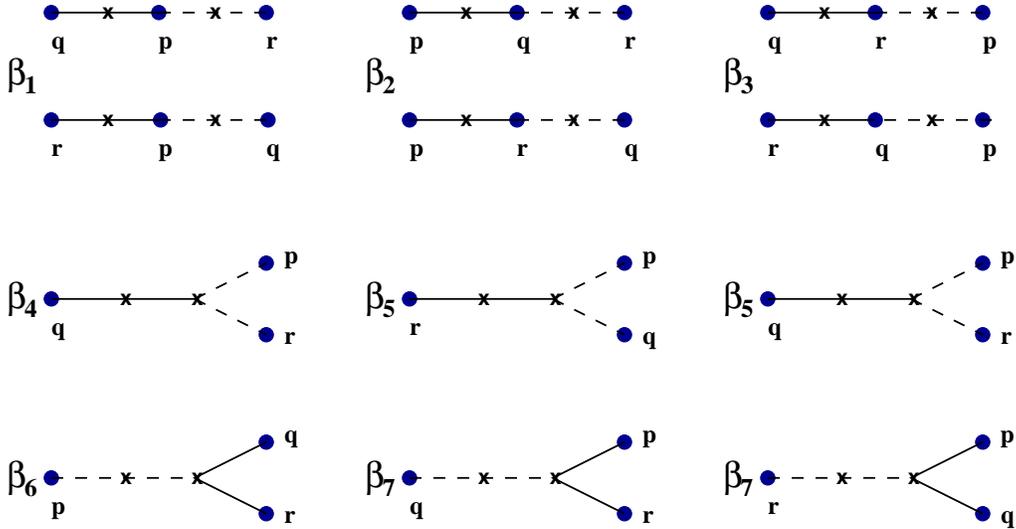} \vspace{-17mm}
\caption{Representation of the {\it ansatz} $\tilde F^{(2)}_{{\bf
p}({\bf q} {\bf r})}$ in (\ref{ea7}) in terms of tree graphs. The
notation used in this figure is the same as that used in
Fig.~\ref{f2}. The crosses in the graphs are vertices at which no
momentum is flowing into or out of the graph.} \label{f3}
\end{figure}

The function $\tilde F^{(1)}$ exactly solves (\ref{ea2}) if
\begin{eqnarray}
\alpha_1+\alpha_2+\alpha_3=0, \label{ea5}
\end{eqnarray}
and $\tilde F^{(1)}$ exactly reproduces the right side of
(\ref{ea3}) if
\begin{eqnarray}
\alpha_3-2\alpha_2=2. \label{ea6}
\end{eqnarray}
However, this choice for $\tilde F^{(1)}$ also creates new terms
on the right side of (\ref{ea3}) that must be eliminated. To
eliminate these additional terms we add to $\tilde F^{(1)}$ a new
{\it ansatz}, designated $\tilde F^{(2)}$. The form of $\tilde
F^{(2)}$ is
\begin{eqnarray}
{\tilde F}^{(2)}_{{\bf p}({\bf q}{\bf
r})}&=&\beta_1\left(\frac{{\tilde G}_{\bf q}^2}{{\tilde G}_{\bf
r}^2}+\frac{{\tilde G}_{\bf r}^2}{{\tilde G}_{\bf q}^2}
\right)+\beta_2\left(\frac{{\tilde G}_{\bf p}^2}{{\tilde G}_{\bf
q}^2}+\frac{{ \tilde G}_{\bf p}^2}{{\tilde G}_{\bf
r}^2}\right)+\beta_3\left(\frac{{\tilde G}_{\bf q}^2}{{\tilde
G}_{\bf p}^2}+\frac{{\tilde G}_{\bf r}^2}{{\tilde G}_{\bf
p}^2}\right)+\beta_4\frac{{\tilde G}_{\bf p}^2}{{\tilde G}_{\bf
q}{\tilde G}_{
\bf r}}\nonumber\\
&&+\beta_5\left(\frac{{\tilde G}_{\bf q}^2}{{\tilde G}_{\bf
p}{\tilde G}_{\bf r}}+\frac{{\tilde G}_{\bf r}^2}{{\tilde G}_{\bf
p}{\tilde G}_{\bf q}}\right)+ \beta_6\frac{{\tilde G}_{\bf
q}{\tilde G}_{\bf r}}{{\tilde G}_{\bf p}^2}+\beta_7
\left(\frac{{\tilde G}_{\bf p}{\tilde G}_{\bf q}}{{\tilde G}_{\bf
r}^2}+\frac{{ \tilde G}_p{\tilde G}_{\bf r}}{{\tilde G}_{\bf
q}^2}\right), \label{ea7}
\end{eqnarray}
where the coefficients $\beta_1$, $\beta_2$, $\ldots$, $\beta_7$
must be determined. The twelve terms in (\ref{ea7}) are
represented graphically in Fig.~\ref{f3}.

To determine the coefficients in $\tilde F^{(2)}$ we require that
the right side of (\ref{ea2}) vanish. This gives the three
conditions
\begin{eqnarray}
\beta_1+\beta_2+\beta_3&=&0,\nonumber\\
\beta_4+2\beta_5&=&0,\nonumber\\
\beta_6+2\beta_7&=&0. \label{ea8}
\end{eqnarray}
In addition, we require that the new terms arising on the right
side of (\ref{ea3}) as a result of the {\it ansatz} $\tilde
F^{(1)}$ must be eliminated. This requirement gives the equations
\begin{eqnarray}
2\alpha_3-\alpha_1-\beta_5&=&0,\nonumber\\
\alpha_1-\beta_2+\beta_3-\beta_7&=&0,\nonumber\\
\alpha_2-\beta_2+\beta_6-\beta_7&=&0,\nonumber\\
\alpha_1-\alpha_2+\beta_3-\beta_6&=&0. \label{ea9}
\end{eqnarray}
One of these equations is redundant because if we subtract the
third equation from the second, we obtain the fourth.

\begin{figure}[t]
\vspace{4.95in} \includegraphics{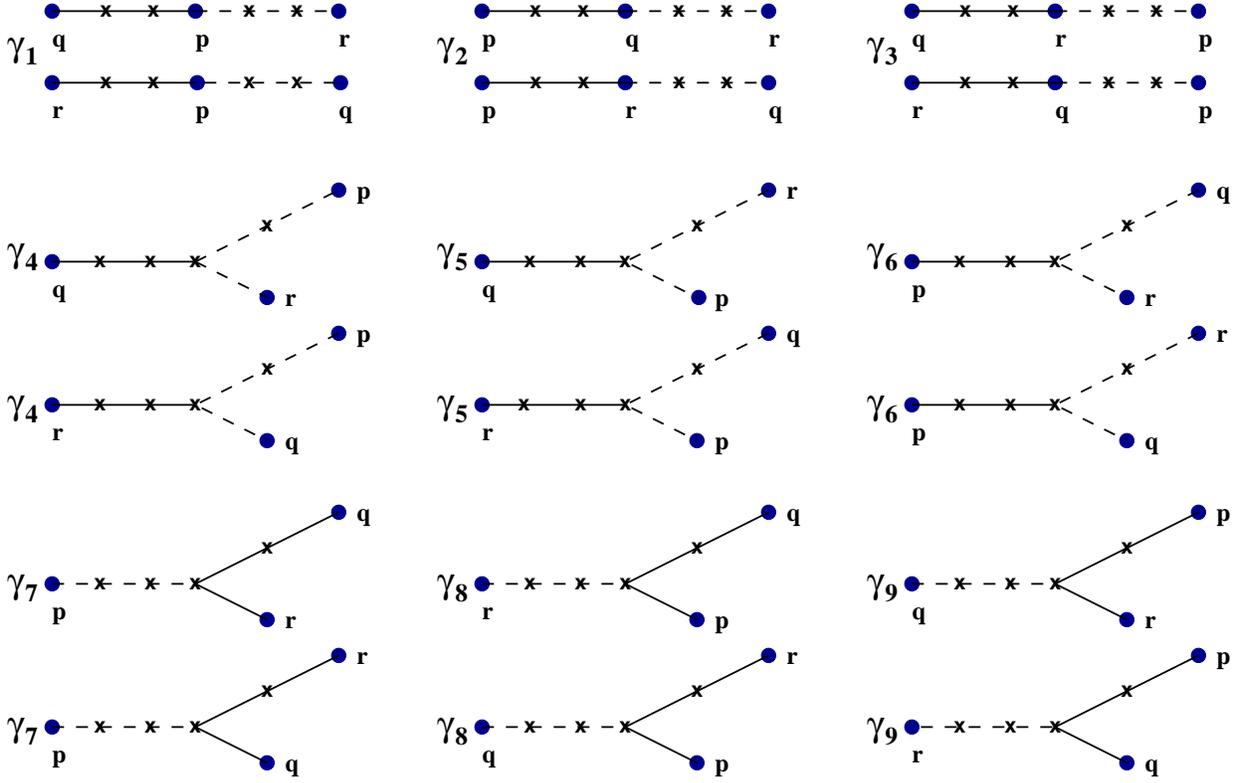} \vspace{-17mm}
\caption{Representation of the {\it ansatz} $\tilde F^{(3)}_{{\bf
p} ({\bf q}{\bf r})}$ in (\ref{ea10}) in terms of tree graphs. The
notation used in this figure is the same as that used in
Fig.~\ref{f3}.} \label{f4}
\end{figure}

We now continue the process. It is necessary to introduce a third
{\it ansatz} in order to eliminate the new terms that appear on
the right side of (\ref{ea3}) as a consequence of the {\it ansatz}
$\tilde F^{(2)}$. This new {\it ansatz} is given by
\begin{eqnarray}
\tilde F^{(3)}_{{\bf p}({\bf q}{\bf
r})}&=&\gamma_1\left(\frac{\tilde G_{\bf q} ^3}{\tilde G_{\bf
r}^3}+\frac{{\tilde G}_{\bf r}^3}{{\tilde G}_{\bf q}^3}\right)
+\gamma_2\left(\frac{{\tilde G}_{\bf p}^3}{{\tilde G}_{\bf
q}^3}+\frac{{\tilde G }_{\bf p}^3}{{\tilde G}_{\bf
r}^3}\right)+\gamma_3\left(\frac{{\tilde G}_{\bf q} ^3}{{\tilde
G}_{\bf p}^3}+\frac{{\tilde G}_{\bf r}^3}{{\tilde G}_{\bf p}^3}
\right)\nonumber \\
&&+\gamma_4\left(\frac{{\tilde G}_{\bf q}^3}{{\tilde G}_{\bf
p}^2{\tilde G}_{\bf r}}+\frac{{\tilde G}_{\bf r}^3}{{\tilde
G}_{\bf p}^2{\tilde G}_{\bf q}}\right)+
\gamma_5\left(\frac{{\tilde G}_{\bf q}^3}{{\tilde G}_{\bf
p}{\tilde G}_{\bf r}^2 }+\frac{{\tilde G}_{\bf r}^3}{{\tilde
G}_{\bf p}{\tilde G}_{\bf q}^2}\right)+
\gamma_6\left(\frac{{\tilde G}_{\bf p}^3}{{\tilde G}_{\bf
r}{\tilde G}_{\bf q}^2 }+\frac{{\tilde G}_{\bf p}^3}{{\tilde
G}_{\bf q}{\tilde G}_{\bf r}^2}\right)
\nonumber\\
&&+\gamma_7\left(\frac{{\tilde G}_{\bf q}{\tilde G}_{\bf
r}^2}{{\tilde G}_{\bf p}^3}+\frac{{\tilde G}_{\bf r}{\tilde
G}_{\bf q}^2}{{\tilde G}_{\bf p}^3}\right)
+\gamma_8\left(\frac{{\tilde G}_{\bf p}{\tilde G}_{\bf
r}^2}{{\tilde G}_{\bf q} ^3}+\frac{{\tilde G}_{\bf p}{\tilde
G}_{\bf q}^2}{{\tilde G}_{\bf r}^3}\right)+
\gamma_9\left(\frac{{\tilde G}_{\bf p}^2{\tilde G}_{\bf
r}}{{\tilde G}_{\bf q}^3 }+\frac{{\tilde G}_{\bf p}^2{\tilde
G}_{\bf q}}{{\tilde G}_{\bf r}^3}\right) \label{ea10}
\end{eqnarray}
and it is represented graphically in Fig.~\ref{f4}.

To determine the nine numerical coefficients in $\tilde F^{(3)}$
we require that the right side of (\ref{ea2}) vanish. This gives
the three conditions
\begin{eqnarray}
\gamma_1+\gamma_2+\gamma_3&=&0,\nonumber\\
\gamma_4+\gamma_5+\gamma_6&=&0,\nonumber\\
\gamma_7+\gamma_8+\gamma_9&=&0. \label{ea11}
\end{eqnarray}
Also, we require that the new terms arising on the right side of
(\ref{ea3}) as a result of the {\it ansatz} $\tilde F^{(2)}$ must
be eliminated. This requirement gives the equations
\begin{eqnarray}
\beta_1-\beta_4+\gamma_5-\gamma_6&=&0,\nonumber\\
\beta_3-\beta_4+\beta_5-\gamma_6&=&0,\nonumber\\
\beta_3+\beta_5-\beta_1-\gamma_5&=&0,\nonumber\\
\beta_1-\beta_2+\gamma_3-\gamma_7&=&0,\nonumber\\
\beta_1-\gamma_2+\gamma_3-\gamma_9&=&0,\nonumber\\
\beta_2-\gamma_2+\gamma_7-\gamma_9&=&0,\nonumber\\
\beta_7+\gamma_7-2\gamma_8&=&0. \label{ea12}
\end{eqnarray}
Note that two of these equations are redundant; if we subtract the
first from the second we obtain the third, and if we subtract the
fourth from the fifth we obtain the sixth.

\begin{figure}[t]
\vspace{4.25in} \includegraphics{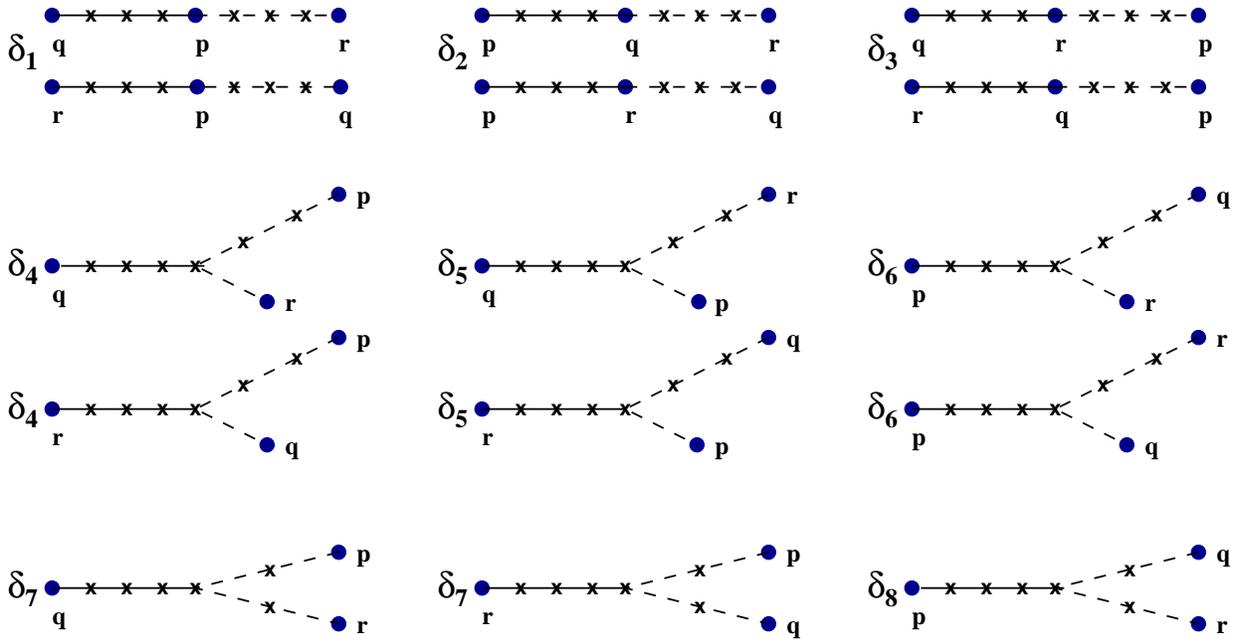} \vspace{-17mm}
\caption{Representation of the {\it ansatz} $\tilde F^{(4)}_{{\bf
p}({\bf q}{\bf r})}$ in (\ref{ea13}) in terms of tree graphs. The
notation used in this figure is the same as that used in
Fig.~\ref{f3}. Note that the number of linear equations that
determine the numerical coefficients is less than the number of
coefficients. This allows us the freedom to exclude {\it a priori}
all graphs having split legs consisting of solid lines.}
\label{f5}
\end{figure}

It is clear that if we continue this process, we obtain more and
more linear equations to solve. However, with each new {\it
ansatz} the number of unknowns always exceeds the number of
equations by one. For example, there are three $\alpha$'s but only
two equations (\ref{ea5}) and (\ref{ea6}); there are seven
$\beta$'s but only six independent equations (\ref{ea8}) and
(\ref{ea9}); there are nine $\gamma$'s but only eight independent
equations (\ref{ea11}) and (\ref{ea12}). Hence, with each new {\it
ansatz} we have a kind of gauge freedom and we find it convenient
to use this freedom to eliminate all graphs having split legs
consisting of solid lines. That is, we choose $\beta_6=\beta_7=0$,
$\gamma_7=\gamma_8=\gamma_9=0$, and so on. We can also choose
$\alpha_1=0$. Thus, for the next iteration our {\it ansatz} reads
\begin{eqnarray}
{\tilde F}^{(4)}_{{\bf p}({\bf q}{\bf r})}&=&\delta_1\left(
\frac{{\tilde G}_{ \bf q}^4}{{\tilde G}_{\bf r}^4}+\frac{{\tilde
G}_{\bf r}^4}{{\tilde G}_{\bf q}^4
}\right)+\delta_2\left(\frac{{\tilde G}_{\bf p}^4}{{\tilde G}_{\bf
q}^4}+\frac{ {\tilde G}_{\bf p}^4}{{\tilde G}_{\bf
r}^4}\right)+\delta_3\left(\frac{{\tilde G }_{\bf q}^4}{{\tilde
G}_{\bf p}^4}+\frac{{\tilde G}_{\bf r}^4}{{\tilde G}_{\bf p
}^4}\right)\nonumber\\
&&+\delta_4\left(\frac{{\tilde G}_{\bf q}^4}{{\tilde G}_{\bf
p}^3{\tilde G}_{\bf r}}+\frac{{\tilde G}_{\bf r}^4}{{\tilde
G}_{\bf p}^3{\tilde G}_{\bf q}}\right)+
\delta_5\left(\frac{{\tilde G}_{\bf q}^4}{{\tilde G}_{\bf
p}{\tilde G}_{\bf r}^3 }+\frac{{\tilde G}_{\bf r}^4}{{\tilde
G}_{\bf p}{\tilde G}_{\bf q}^3}\right)+
\delta_6\left(\frac{{\tilde G}_{\bf p}^4}{{\tilde G}_{\bf
r}{\tilde G}_{\bf q}^3 }+\frac{{\tilde G}_{\bf p}^4}{{\tilde
G}_{\bf q}{\tilde G}_{\bf r}^3}\right)
\nonumber\\
&&+\delta_7\left( \frac{{\tilde G}_{\bf q}^4}{{\tilde G}_{\bf
p}^2{\tilde G}_{ \bf r}^2}+\frac{{\tilde G}_{\bf r}^4}{{\tilde
G}_{\bf p}^2{\tilde G}_{\bf q}^2} \right)+\delta_8\frac{{\tilde
G}_{\bf p}^4}{{\tilde G}_{\bf q}^2{\tilde G}_{\bf r}^2}.
\label{ea13}
\end{eqnarray}
The graphical representation of $\tilde F^{(4)}$ is shown in
Fig.~\ref{f5}.

Requiring that the right side of (\ref{ea2}) continue to vanish
gives the three conditions
\begin{eqnarray}
\delta_1+\delta_2+\delta_3&=&0,\nonumber\\
\delta_4+\delta_5+\delta_6&=&0,\nonumber\\
2\delta_7+\delta_8&=&0. \label{ea14}
\end{eqnarray}
Also, to eliminate the new terms arising on the right side of
(\ref{ea3}) as a result of the {\it ansatz} $\tilde F^{(3)}$ we
must require that
\begin{eqnarray}
\gamma_2-\gamma_6-\delta_4+\delta_5&=&0,\nonumber\\
\gamma_3+\gamma_5-\gamma_6-\delta_4&=&0,\nonumber\\
\gamma_3+\gamma_5-\gamma_2-\delta_5&=&0,\nonumber\\
\gamma_6-2\gamma_4+\delta_7&=&0,\nonumber\\
\gamma_1-\gamma_2-\delta_3&=&0,\nonumber\\
\gamma_1-\delta_1&=&0,\nonumber\\
\gamma_2-\delta_1+\delta_3&=&0. \label{ea15}
\end{eqnarray}
Note that two of these equations are redundant; if we subtract the
first from the second we obtain the third ,and if we subtract the
fifth from the sixth we obtain the seventh. Thus, in total there
are eight $\delta$'s to be determined by eight independent
equations.

Let us now examine the solutions for the undetermined numerical
coefficients. Solving the system of linear equations for the first
five groups of these coefficients, we find that
\begin{eqnarray}
&& \alpha_1=0,\quad\alpha_2=-2,\quad\alpha_3=2,\nonumber\\
&&\beta_1=0,\quad\beta_2=-2,\quad\beta_3=2,\quad\beta_4=6,
\quad\beta_5=-12,
\nonumber\\
&& \gamma_1=0,\quad\gamma_2=-2,\quad\gamma_3=2,\quad
\gamma_4=20,\quad\gamma_5
=10,\quad\gamma_6=- 30,\nonumber\\
&& \delta_1=0,\quad\delta_2=-2,\quad\delta_3=2,\quad\delta_4=
42,\quad\delta_5
=14,\quad\delta_6=-56,\quad\delta_7=70,\quad\delta_8=-140,
\nonumber\\
&&\epsilon_1=0,\quad\epsilon_2=-2,\quad\epsilon_3=2,\quad
\epsilon_4=72,\quad
\epsilon_5=18,\quad\epsilon_6=-90,\quad\epsilon_7=252,
\quad\epsilon_8=168,
\nonumber\\
&&\quad\epsilon_9=-420. \label{ea16}
\end{eqnarray}
A brief examination of these coefficients shows that these numbers
are all binomial coefficients. In fact, by inspection we can now
write down explicitly all terms in $\tilde F^{(1)}$, $\tilde
F^{(2)}$, $\tilde F^{(3)}$, and so on, as the following double
sum:
\begin{eqnarray}
{\tilde F}_{{\bf p}({\bf q}{\bf r})} = -4\sum_{n=1}^\infty
\frac{{\tilde G}_{\bf p}^n}{{\tilde G}_{\bf r}^n} \sum_{k=0}^n
\left({2n}\atop{2k}\right) \frac{{\tilde G}_{\bf r}^k} {{\tilde
G}_{\bf q}^k} +4\sum_{n=0}^\infty \frac{{\tilde G}_{\bf
r}^{n+1}}{{\tilde G}_{\bf p}^{n+1}} \sum_{k=0}^n
\left({2n+1}\atop{2n+1-2k}\right)\frac{{\tilde G}_{\bf p}^k}
{{\tilde G}_{\bf q}^k}. \label{ea17}
\end{eqnarray}
It is easy to perform these sums, and the result for $\tilde N$ is
given in (\ref{e80}). It is remarkable that the solution for
$\tilde N$ is {\it unique} despite the arbitrary choice of
``gauge'' that we have made in solving the systems of algebraic
equations.

We emphasize that all of the tree graphs that appear in this
analysis have three external lines. This reflects the fact that
the form factors $M$ and $N$ that we are calculating represent the
contribution of tree graphs to the three-point vertices. The next
order in perturbation theory is proportional to $\epsilon^3$, and
the graphical representation of the result consists of all tree
graphs having five external legs.

\end{document}